\documentclass{emulateapj}

\usepackage{lscape}

\def\degs#1.#2 {$#1{}\rlap{\rm .}^\circ#2$}            
\def\by {$\times\,$}

\bibpunct{(}{)}{;}{a}{}{,}

\newcommand{\asca}{{\emph{ASCA}}}
\newcommand{\chandra}{{\emph{Chandra}}}
\newcommand{\conx}{{\emph{Constellation-X}}}
\newcommand{\xmm}{\emph{XMM-Newton}}
\newcommand{\sax}{{\emph{Beppo-SAX}}}

\newcommand{\hst}{{\emph{HST}}}

\newcommand{\feka}{\mbox{Fe\,K$\alpha$}}
\newcommand{\kms}{\mbox{\,km\,s$^{-1}$}}
\newcommand{\cmsq}{\mbox{\,cm$^{-2}$}}

\newcommand{\lumin}{\mbox{\,ergs~s$^{-1}$}}
\newcommand{\persec}{\mbox{\,s$^{-1}$}}
\newcommand{\nh}{\mbox{${N}_{\rm H}$}}

\newcommand{\CIV}{\ion{C}{4}}

\newcommand{\OIII}{[\ion{O}{3}]}
\newcommand{\Halpha}{H$\alpha$}

\newcommand{\mrk}{{Mrk~231}}

\slugcomment{To Appear in The Astrophysical Journal}

\shortauthors{Gallagher et al.}
\shorttitle{Variability Study of Mrk 231}

\begin{document}


\title{Variation in the Scattering Shroud Surrounding Markarian 231\altaffilmark{1,2}}
\author{
S. C. Gallagher,\altaffilmark{3}
Gary D. Schmidt,\altaffilmark{4}
Paul S. Smith,\altaffilmark{4}
W. N. Brandt,\altaffilmark{5}
G. Chartas,\altaffilmark{5} \\
Shavonne Hylton,\altaffilmark{6}
D. C. Hines,\altaffilmark{7}
and  M. S. Brotherton\altaffilmark{8} 
}

\altaffiltext{1}{Some of the data presented here were obtained with the MMT
Observatory, a facility operated jointly by The University of Arizona and the
Smithsonian Institution.}

\altaffiltext{2}{Based on observations obtained with the Upgraded
  Fiber Optic Echelle Spectrograph on the Hobby-Eberly Telescope, 
which is operated by McDonald Observatory on behalf of the University
  of Texas at Austin, the Pennsylvania State University, Stanford
  University, the Ludwig-Maximillians-Universit\"at, Munich, and the
  George-August-Universit\"at, G\"ottingen.}

\altaffiltext{3}{Department of Physics \& Astronomy, University of
  California -- Los Angeles, Mail Code 154705, 475 Portola Plaza, Los
  Angeles CA, 90095--4705, sgall@astro.ucla.edu}

\altaffiltext{4}{Steward Observatory, The University of Arizona, Tucson AZ
85721; gschmidt@as.arizona.edu, psmith@as.arizona.edu}

\altaffiltext{5}{Department of Astronomy \& Astrophysics, The
  Pennsylvania State University, University Park, PA 16802;
  niel@astro.psu.edu, chartas@astro.psu.edu}

\altaffiltext{6}{Center for Space Research, Massachusetts Institute of
  Technology, 77 Massachusetts Avenue, Cambridge, MA 02139}

\altaffiltext{7}{Space Science Institute, 4750 Walnut Street, Suite
  205, Boulder, CO 80301; dean.hines@colorado.edu}

\altaffiltext{8}{Department of Physics \& Astronomy, University of
  Wyoming, Dept. 3905, Laramie, WY 82071; mbrother@uwyo.edu}

\begin{abstract}
We present a detailed study of the nuclear structure of
the highly polarized broad absorption line quasar, \mrk, through a
multiwavelength campaign of \chandra\ observations, optical
spectroscopy, optical spectropolarimetry, and imaging polarimetry.
This campaign was designed to extend the 40~ks \chandra\ study of Gallagher et
al. and the optical and UV spectropolarimetric study of Smith et
al. to probe variability on multiple timescales.   
As direct emission from the nucleus is heavily obscured at optical
through X-ray wavelengths, the detailed study of scattered emission has
lead to insights into the stratified and complex central region of
this active galaxy.  
Though significant continuum variability is not detected in any of the
three new 40~ks \chandra\ observations, we investigate \feka\
emission features in the individual and combined spectra.  Comparing
echelle spectra of the \ion{Na}{1}\,D absorption lines with the
literature, we show that one system disappeared in 2000 only to
reappear in later epochs. 
Notably, we detect a large decrease in polarization across the
entire optical spectrum of \mrk\ from 1995 to 2003.  Though the
polarization fraction fell, e.g., from 4\% to 3\% at $\lambda_{\rm
  obs}\sim6000$\AA, the polarization position angle spectrum remained
unchanged.  The optical polarization behavior is consistent with a
decrease in the flux scattered by circumnuclear material on spatial
scales where the broad-line region is resolved by the scattering
material.  Ultraviolet imaging
polarimetry of \mrk\ by the {\em Hubble Space Telescope} sets an upper
limit on the distance between the active nucleus and the scattering
regions of $<$20\,pc.
\end{abstract}


\section{Introduction}

The variability of emission from active galactic nuclei (AGNs) has long been
employed in efforts to discern the structure and physics of material near the
central energy source.  The dramatic changes in brightness and
polarization of the continua of radio-loud objects, particularly at short
wavelengths, played an important role in developing the popular model of a relativistically beamed jet
emitting by the synchrotron mechanism \citep[e.g.,][]{br1978}.  The effects
of more isotropic changes in luminosity have been exploited by the
``reverberation mapping'' technique to infer the size, thickness, and
ionization properties of the broad emission-line region, and more recently to
identify systematic trends in nuclear parameters between various classes of
AGNs \citep[e.g.,][]{NePe97}.

Evidence for changes of the {\em structure} of the nuclei of active galaxies
can be found in the appearance and disappearance of shoulders and other features
in the profiles of broad emission lines, modifications of the Balmer
decrements, and in certain cases in the replacement of a normally
centrally peaked profile by one that is strongly double-peaked
(\citealt{UlMaUr97} and references therein).  Some of these effects appear
to be induced by luminosity changes of the central continuum source, but
others require mass motions or the redirection of an illuminating
beam, also known as the `searchlight' effect.  At
larger radii, variations in the properties of absorbing gas
as indicated by changes in UV absorption lines have been detected on timescales of
$\sim$1 yr in several Seyfert galaxies, with explanations including both
luminosity-induced ionization changes and the motion of absorbers across our
line of sight (e.g., \citealt{gabel+03} and references therein).  
Similarly, large column density variations have been observed in X-ray
studies of absorbed AGNs on timescales of months to years
\citep[e.g.,][]{Ris+02,gall04}. 

 The luminous Type~1 (broad emission line) AGN Markarian~231 is well
known to exhibit no less than three displaced optical absorption-line systems
in \ion{Na}{1}, \ion{He}{1}, and \ion{Ca}{2} \citep{BoEtal1977}, and a
fourth appeared in observations made in 1988 \citep{BoMeMoPe1991}.  Further
variability in the strengths of the absorption lines has been reported in
subsequent years \citep[][hereafter FRM]{KoDiHa1992,FoRiMc1995}
and \ion{Mg}{2}$\lambda$2800 absorption
associated with one of the systems reported by 
\citet[][hereafter SSAA]{Smith+95}.  
This observation, coupled with the source's high intrinsic luminosity, 
qualified \mrk\ as a low-ionization broad absorption-line (LoBAL) quasar.
LoBAL quasars are notoriously weak X-ray emitters
\citep{green+01,gall+02}; therefore the
proximity ($z=0.0422$) of \mrk\ makes it by far the most easily studied
example of this class in X-rays.

The fact that most of the $\sim$$4\times10^{12}$~$L_\sun$ luminosity of
\mrk\ appears in the infrared \citep{RiLo1972} was early evidence that the
central energy source is heavily obscured along most sightlines.  This
is strongly supported by the paucity of emission from the narrow-line
region \citep[specifically, the absence of \OIII\ flux;][]{HaKe1987},
and the shape and luminosity of the 2--10~keV X-ray spectrum imply that the
nucleus is only seen via reflection and/or scattering in that
band \citep[][hereafter G02]{MaRe2000,gall_mrk}.  An analysis of 15--60~keV \sax\ Phoswich Detector
System (PDS) data by \citet{braito+04} confirmed this interpretation
with the detection, for the first time, of the direct X-ray continuum
absorbed by an intrinsic column density of \mbox{\nh$\sim 2\times10^{24}$\cmsq}.
Absorption of this magnitude completely blocks the direct 2--10~keV continuum,
which has an absorption-corrected luminosity $L_{\rm
  2-10}\sim10^{44}$\lumin\ \citep{braito+04}.  The observed, nonthermal 2--8~keV
continuum, with $L_{\rm X}\sim2\times10^{42}$\lumin, must therefore be
reprocessed emission. However, the 2--8~keV flux variability on timescales of hours
detected by G02 with \chandra\  indicates that the
X-ray scattering medium is compact, with spatial scale of $\sim10^{15}$\,cm.
Given the proximity of this gas to the central continuum source,
this scattering medium is likely to be highly ionized.
Obscuration and the attendant scattering of nuclear light by pervasive dust clouds at
much larger radii, $R\gtrsim1$~pc, give rise to polarization of
the optical-UV light,
and polarization fractions that reach 15$\%$ in the near-UV have made the object a popular
target for polarimetric studies with increasing quality over the past two
decades (\citealt{Thompson+80,SchMil85,GoMi1994}; SSAA; \citealt{Smith+04}).

The sensitivity of both the X-ray spectrum and the optical
polarization of \mrk\ to the
geometry of scattering on different physical scales, and the insight that
variations in these diagnostics
might lend to our knowledge of the structure of the nuclear regions of a
luminous Type~1 AGN, have led us to carry out a program of
high-quality, ground-based optical spectropolarimetry and 
0.5$-$8 keV spectroscopy with the {\em Chandra} X-ray
Observatory \citep{chandra_ref}.
In this paper we present the results of that study as well as 
optical echelle spectroscopy and UV polarimetry from different epochs that
contribute additional information on the properties of the nucleus. 

The cosmology assumed throughout has $H_0=70$~km~s$^{-1}$~Mpc$^{-1}$,
$\Omega_{\rm M}=0.3$, and $\Omega_{\Lambda}=0.7$.  For \mrk\ at $z=0.0422$, 1
arcsecond corresponds to a physical scale of 833~pc.

\ \ \ \ ~
\ \ \ \ ~

\ \ \ \ ~
\ \ \ \ ~

\section{X-ray Observations and Analysis}
\label{sec:xray_data}

\mrk\ has been observed 4 times with \chandra, each time at the
aimpoint of the ACIS \citep{acis_ref} back-illuminated S3
detector. The first observation of 2000 Oct 19 (observation 1031) was
in full-frame mode.  The subsequent three observations of 2003 Feb
(observations 4028--30) were
performed in 1/2 subarray mode in order to reduce the $\sim7\%$ pile-up
fraction of the first observation (G02).  Each
ACIS observation was between 36--39~ks long; Table~\ref{tab:xlog} 
presents the log of the observations, including dates, exposure times,
and soft and hard-band count rates.  The $\Delta t$ of
$\sim$1\,week between observations 4028, 4029, and 4030 was chosen to probe times with a
scale intermediate between the hours and years time scales of previous
observations.  Physically, this corresponds to structure on scales of
$\sim10^{16}$\,cm, the scale of the thick, highly ionized gas postulated by
 \citet{MuChGrVo1995} to shield a radiatively driven BAL wind from
 becoming overionized by the ionizing X-ray continuum.

The \chandra\ data were processed with the standard \chandra\ X-ray Center
(CXC) pipeline from which the level 1 events file was used.
 The data were filtered on good time intervals.
The standard pipeline processing introduces a $0\farcs5$ software randomization into the
event positions, which degrades the spatial resolution by $\sim12\%$
\citep{ChartasEtal2002}. This randomization is done to avoid aliasing
effects noticeable in observations with exposure times of less than
2~ks. Since spatial resolution was
important for our analysis and the ACIS observations were significantly
longer than 2~ks, we removed this randomization from the ACIS
observation by recalculating the event positions in
sky coordinates using the CIAO tool, {\em acis\_process\_events}. 
The ACIS-S3 imaging data were corrected for radiation damage sustained by the
CCDs during the early months of \chandra\ operations using the
software described in \citet{cti_ref}.  This procedure mitigates
the effects of charge transfer inefficiency (CTI)
that introduces positional dependence of the gain, quantum
efficiency, and grade distribution. CTI-corrected ACIS-S3 data
allow a single redistribution matrix file (rmf) to be used across the whole CCD.
After CTI-correction, the data were filtered to keep ``good'' time
intervals and \asca\ grades (grades 0, 2, 3, 4, and 6). 
Filtering on ``status'' can remove X-ray events from the nucleus
because X-rays can be inappropriately tagged as cosmic-ray
afterglows. Therefore, the values in the status column for cosmic-ray
afterglows were ignored for the final filter on ``good'' status. 

\subsection{Variability Analysis}

Given the significant hard-band X-ray variability detected in the
first \chandra\ observation, we first searched for
count-rate changes both within and between the observations. 
The time and energy of each event within a
3$\arcsec$-radius source circle centered on the peak of the X-ray
emission were extracted from the reduced events file for each observation.  
The unbinned full (0.5--8.0~keV), soft
(0.5--2.0~keV), medium (2.0--5.0~keV),
hard (2.0--8.0~keV), and very hard (5.0--8.0~keV) band event arrival
times for each observation were evaluated with the non-parametric 
Kolmogorov-Smirnov test to search for variability within each
observation.  For the new observations, no significant variability was
detected in any band. The original data  from 2000 (observation 1031) 
were downloaded from the public archive, rereduced with CIAO 3.0 as
outlined above, and retested. As presented in G02, we recovered
significant variability in the medium and hard bands at
the $>99\%$ confidence level; this result was also confirmed by the 
independent analysis of \citet{ptak+03}.

\begin{figure}
\vspace{3.7in}
\includegraphics{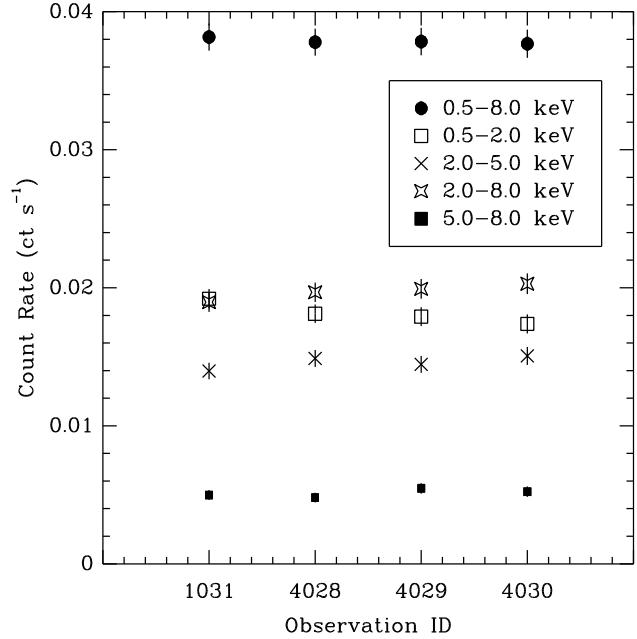}
\caption{
Count rates for the four different \chandra\ observations of \mrk. The
decrease in the 0.5--2.0~keV count rate from observation 1031 to 4028
is likely a result of the declining soft-band effective area over
time.   Though the full-band count rate has remained
remarkably stable, the slight changes in soft and hard-band count
rates suggest a hardening of the spectrum.}
\label{fig:cr}
\end{figure}

To check for variability between observations, the background-subtracted
nuclear count rates were compared in the five bands. The chosen background region was a source-free annulus with an inner radius of
$40\arcsec$ and an outer radius of $50\arcsec$, well outside of the
galaxy, centered on the nucleus.  Figure~\ref{fig:cr} presents these
data.  The full-band count rate is remarkably stable, but there
appears to be a slight shift over the course of observations, with
more hard-band counts relative to soft-band counts 
in the last observation than in the preceeding ones. This trend is
slight however, and we investigate any possible spectral hardening
further with spectral analysis.

\subsection{Nuclear Spectroscopy}
The nuclear spectra were extracted from each reduced \chandra\ dataset
independently.  The source and background regions were the same as
those used in the variability analysis.
The auxiliary response files (arfs) and rmfs were generated with the CIAO~3.0 tools
{\em mkarf} and {\em mkrmf}, respectively. 

The final, preferred model from G02 for the fit to the \chandra\ spectrum included a
  thermal component from the nuclear starburst as well as three absorbed 
  power-law components with different normalizations and intrinsic
  column densities (see Figure~9d of G02). Given the observed \mbox{2--8~keV}
  luminosity of the source ($\sim100$ times fainter than expected
  based on both the dereddened UV luminosity and the $>10$~keV X-ray
  spectrum) and the known complexity of the scattering
  geometry derived from optical polarization studies (SSAA), these power-law continua
  were interpreted as direct continuum X-rays scattered into the line
  of sight by distinct structures, and therefore passing through
  different amounts of absorption. The scattering was inferred
  to be by electrons in highly ionized plasma, 
thus preserving the shape of the intrinsic spectrum.
  In G02, the photon index of each power-law
  model was fixed to $\Gamma=2.1$, based on the correlation of
  H$\beta$ FWHM and $\Gamma$ found by \citet{ReTu2000} from \asca\ spectral fitting of quasars.  
  To evaluate this model in the context of the
  new \chandra\ data, we examined the combined, average spectrum from
  the three new observations to increase the signal-to-noise ratio.
  With these data, the power-law photon index can be fit independently.
  We do not incorporate the data from the first \chandra\ epoch
  because they were mildly piled-up and taken in a different mode
  (full array rather than 1/2 subarray).

To create an average nuclear spectrum, the three unbinned spectra and
background spectra were each added together using the FTOOL {\em mathpha}.
An average arf and rmf were created by weighting by the 0.5--8.0~keV
counts using {\em addarf} and {\em addrmf}, respectively.  The
combined spectrum (with a total exposure of 114.3~ks and 4300 counts)
was then fit using the X-ray spectral-fitting tool, XSPEC
\citep{xspec_ref}.  Three absorbed power-law models
plus a Raymond-Smith plasma were applied to the combined spectrum;
$\Gamma$ was tied for each power law. An acceptable fit
($\chi^2/\nu=174.8/153$) was found for this model with
$\Gamma=1.92^{+0.37}_{-0.70}$.  However, given the relatively
featureless spectrum, this solution is not unique, and much lower
values of $\Gamma$ also provide statistically acceptable fits.  As
$\Gamma\sim1.9$ is typical of radio-quiet quasars and more plausible
in the context of this model, we fixed $\Gamma=1.92$ to determine the errors in the remaining
parameters. The column densities (assumed to be neutral) for the three
power laws are
\nh=$(0.19^{+0.04}_{-0.03},4.6^{+0.92}_{-0.55},45^{+17}_{-10})\times10^{22}$\cmsq,
similar to the values listed in Table~2 of G02.  This model does
not include a component corresponding to the highly absorbed
(\nh$\sim2\times10^{24}$\cmsq) direct X-ray continuum studied by
\citet{braito+04} because this component does not contribute
significant counts to the \chandra\ spectrum. 
The best-fitting temperature for
the Raymond-Smith plasma is $kT=0.89^{+0.05}_{-0.03}$~keV.  Though
the spectral bump near 1~keV (likely from spectrally 
unresolved Fe~L emission lines) is
fit by this model, it is also possible that there is a contribution
from photoionized plasma, and  \citet[][hereafter TK]{tk03}
suggested that an emission feature seen at
$\sim0.45$~keV in the \xmm\ spectrum might be a \CIV\ radiative
recombination line.  

TK detected a broad, \feka\ emission line
with an equivalent width EW\,$\sim450$~eV, a rest-frame energy
$E=6.49\pm0.12$~keV, and $\sigma=0.17\pm0.12$~keV in the 17.3~ks \xmm\ EPIC-pn
spectrum of Mrk~231. Such a strong, broad feature is not evident in this
0.5--8.0~keV combined spectrum, and adding a broad line does not make a significant
improvement in the spectral fit ($\Delta\chi^2=-2.8$ for three fewer
degrees of freedom).  Adding a single, unresolved line in the Fe region has a best-fitting
energy of $6.64^{+0.05}_{-0.09}$~keV and gives $\Delta\chi^{2}=-5.5$ for two
fewer degrees of freedom.

\begin{figure}
\vspace{3.5in}
\includegraphics{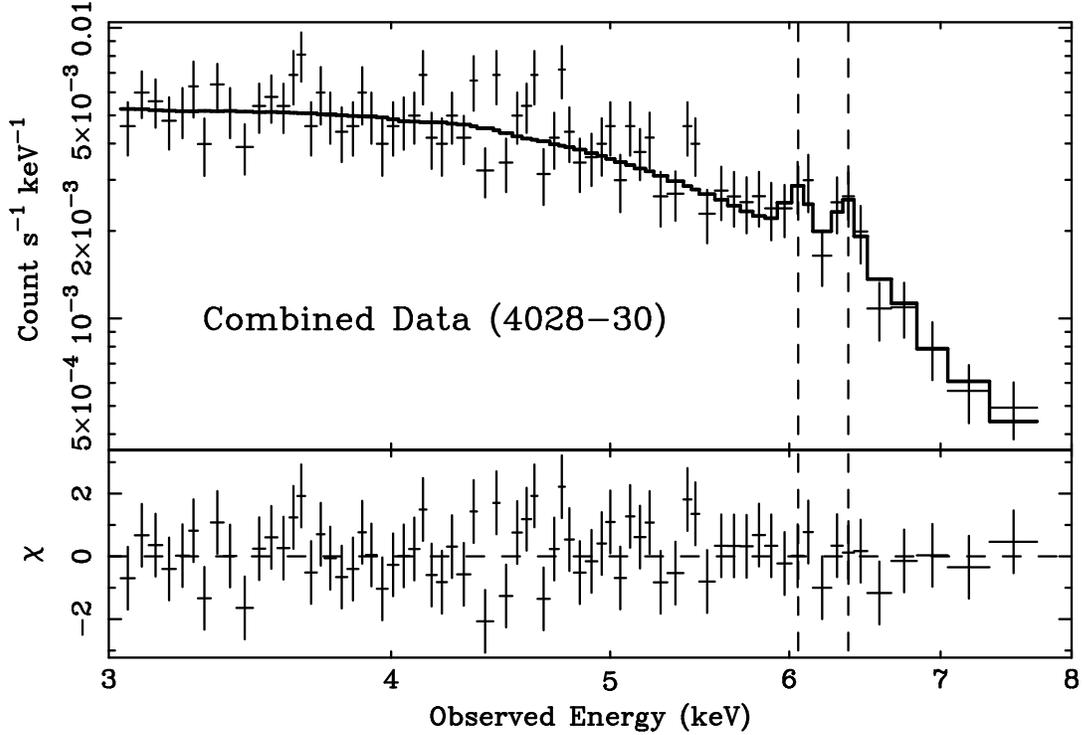}
\caption{
X-ray spectrum and model from the combined 3--8~keV \chandra\ data from
observations 4028--30.  The thick histogram is the best-fitting
power-law model with two narrow gaussian emission lines and an Fe edge
(see Table~\ref{tab:xspec}) convolved with the \chandra\ response. 
Vertical dashed lines mark the best-fitting energies of two narrow
Fe emission lines.
\label{fig:xspec}
}
\end{figure}

To focus further on the nuclear spectrum in general and the \feka\
region in particular, we follow the example of TK by ignoring
the data below 3.0~keV where starburst emission is significant and fitting
the 3.0--8.0~keV counts with a single power-law model for the
continuum.  
From G02, this is also the region of the spectrum that showed significant variability in
observation 1031 of 2000 Oct. 
To account for
the \feka\ edge absorption expected from multiple absorbers (not
otherwise included in this model), we also add an edge with
the rest-frame energy fixed at 7.12~keV, the ionization energy of
neutral Fe. This two component model is sufficient to fit the data, with
$\chi^2/\nu$=69.0/69, and adding a broad line does not
significantly improve the fit ($\Delta\chi^2=-3.2$ for three fewer
degrees of freedom).  Convolving the best-fitting \xmm\ model with the
\chandra\ response and allowing $\Gamma$, the continuum normalization,
and $\tau_{\rm max}$ (the maximum optical depth of the ionization edge) 
to vary while keeping the \feka\ line parameters
fixed does not provide a better fit ($\chi^2/\nu=72.8/69$) than the power-law plus edge model
alone.  Freeing the continuum parameters is
necessary to account for both cross-calibration uncertainties and
potential source variability. A strong negative residual at
rest-frame $\sim$6.45~keV (approximately the peak energy of the \xmm-detected
emission line) suggests that \chandra\ is resolving the
broad line seen by TK into two narrow lines.
At 6~keV, the \xmm\ EPIC-pn has an effective area about 4 times larger
than that of ACIS-S3.\footnote{http://heasarc.gsfc.nasa.gov/docs/xmm/about\_why.html}
Our 114~ks observation is about 6.6 times longer than the \xmm\
observation, and therefore the \chandra\ data have better photon statistics.
Both the \chandra\ and \xmm\ data can be successfully fit with two
unresolved ($\sigma=0.01$\,keV) gaussian
lines, with consistent line energies of $\sim$6.3 and $\sim$6.6~keV. 
In the \xmm\ data, with fewer
counts in the observed-frame 5.8--6.7~keV band ($\sim140$ versus $\sim210$
for the combined \chandra\ spectrum), two gaussians are not
preferred over a single, broad gaussian.  Figure~\ref{fig:xspec} shows
the combined data and preferred spectral model, and the parameters from the
preferred spectral models for both the \chandra\ and \xmm\ data 
are presented in Table~\ref{tab:xspec}.

From visual inspection of a merged 5.8--6.7~keV (rest-frame
6--7~keV) image, a 3$\arcsec$-radius source cell encompasses all of
the counts in this band.  That is, there is no evidence from the
\chandra\ data of significant Fe-band emission coming from the host
galaxy.  Therefore, any discrepancies between the observation
presented in TK and the \chandra\
data do not arise from the different beam sizes of \chandra\ and \xmm.

\begin{figure}
\vspace{4.2in}
\includegraphics{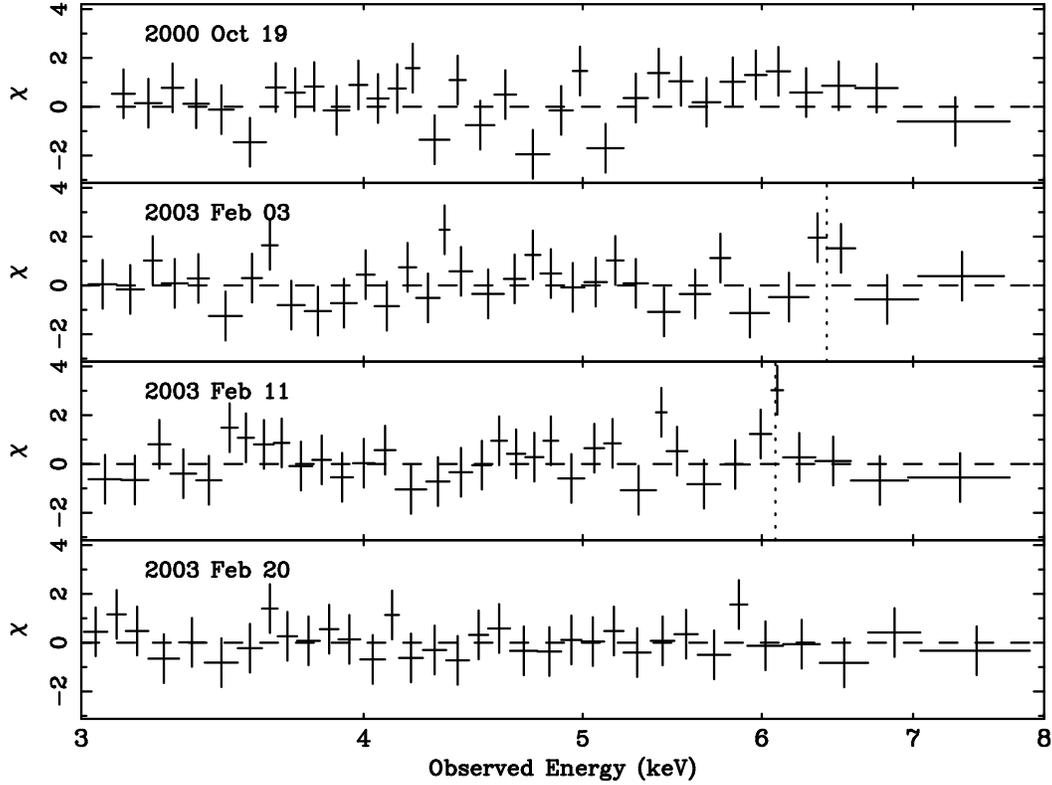}
\caption{Residual plots for the four 3--8~keV \mrk\ spectra from
  joint-spectral fitting. 
  For the joint-fitting, the values of $\Gamma$ and $\tau_{\rm max}$ were
  tied, power law normalizations were free to vary, and the edge
  energy was fixed to 7.12~keV, the energy of the neutral Fe K edge.
  To determine the continuum fit, the iron-band
  region was ignored.
  The energies of the two possible Fe lines are
  marked with vertical dotted lines.}
\label{fig:resid}
\end{figure}

We investigate further the possibility of nuclear spectral variability
between the 5 observations (four \chandra\ and one \xmm).
Again, each spectrum was initially fit jointly
with a power law model and an absorption edge fixed to 7.12~keV.  
Both $\Gamma$ and $\tau_{\rm max}$  were tied for the fit, and the
power-law normalizations were free to vary for each observation.  
To look for iron emission, the data in the iron band
(rest-frame 6--7~keV, observed-frame 5.8--6.7~keV) were ignored for this continuum fit.  The
\chandra\ residuals from this joint-spectral fitting are shown in Figure~\ref{fig:resid}.  
For the final observation, there is no strong evidence of iron emission.
For the first observation, a series of positive residuals in the iron band 
is suggestive of a broad emission feature, as found
by TK. In addition, for the middle two observations, Feb 03 and Feb 11, there are
consecutive positive residuals at the 1--2$\sigma$ level in the
iron-band region. To examine these features more closely, we fit the
spectra individually.

For each observation, the power-law model with an edge at 7.12~keV was
applied to the data.  The values of $\Gamma$, $\tau_{\rm max}$, and the power-law
normalization were free to vary.  In each case, the fit was acceptable
with $\chi^{2}/\nu$=29.8/30, 29.1/31, 30.3/33, and 13.1/31 for
observations 1031, and 4028--30, respectively.  Next, a narrow iron
line was added to each model as a redshifted gaussian with the energy free to vary between
rest-frame 6--7~keV.  The value for $\sigma$ was fixed to 0.01~keV,
unresolved at ACIS CCD resolution.  In chronological order, the new values for
$\chi^{2}$ were 26.8, 22.4, 21.7, and 13.1 for two fewer degrees of
freedom. Of these, only the middle two (observations 4028 and
4029) indicate a significant improvement in the model fit with
$\Delta\chi^2=-6.7$ and --8.6, respectively.  According to the $F$-test,
the addition of the gaussian component improved the individual 
models at the $>97\%$ and $>99\%$ confidence level for the data from observation
4028 and 4029, respectively.  For observation 1031, a broad line with
$\sigma$ free to vary was also tested.  For three fewer degrees of
freedom from the power-law plus edge model, $\Delta\chi^2=-6.6$,
indicating an improvement at the $\sim92$\% level over the continuum
model alone. Formally, these confidence levels should
be lower, as they apply to a single observation (rather
than the four we are presenting), but they are still useful as a
relative gauge of the improvement.  The values and errors for the
parameters of the preferred model for each spectrum are listed in
Table~\ref{tab:xspec}.

\begin{figure}
\vspace{4.0in}
\includegraphics{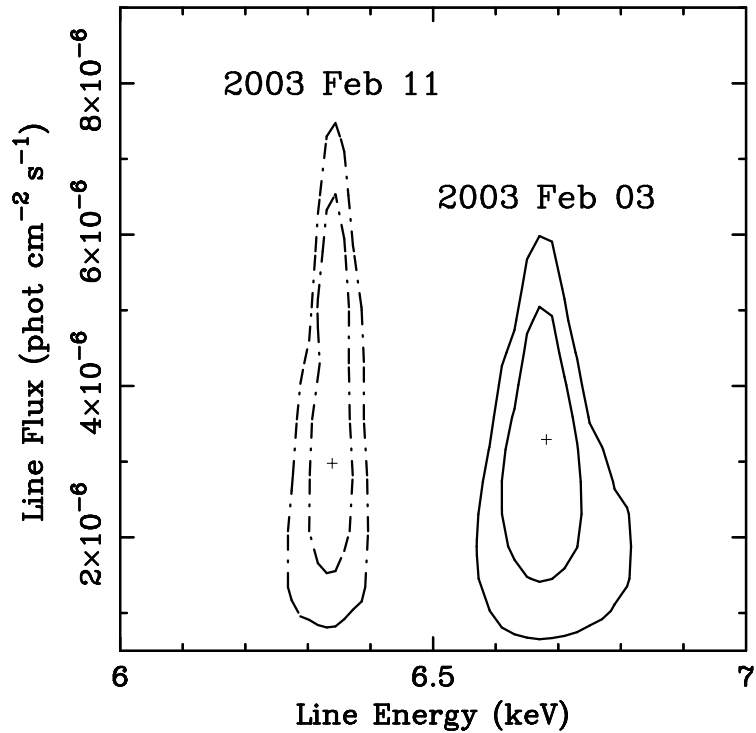}
\caption{
Contour plots of the line flux versus rest-frame energy of the iron
lines detected in the two middle observations.  The contours are $68\%$ and
$90\%$ confidence for two parameters of interest ($\Delta\chi^2=2.31$
and 4.61, respectively).  Though the fluxes and EWs are consistent,
the energies are not.}
\label{fig:cont}
\end{figure}

In Figure~\ref{fig:cont}, the contour plots for the line flux versus
line energy for 4028 and 4029, the observations with two possible narrow lines,
are plotted.   As is evident from the
figure, the $68\%$ and $90\%$ contours do not overlap; the energies
are not consistent.  The fluxes and EWs found for both observations are
however consistent.

\subsection{X-ray Properties of the Galaxy}

A merged event file of the three new observations 
was created using the CIAO script {\em  merge\_all}.  This script
reprojects the positions of events from all images into a common frame
so that they can be combined. The images were then adaptively smoothed
using the CIAO tool {\em csmooth}.  
The X-ray emission from the galaxy based on the 40~ks 1031 observation 
was described in detail by G02, and so we only comment briefly on
the new information obtained from these additional data.  

\begin{figure}
\plottwo{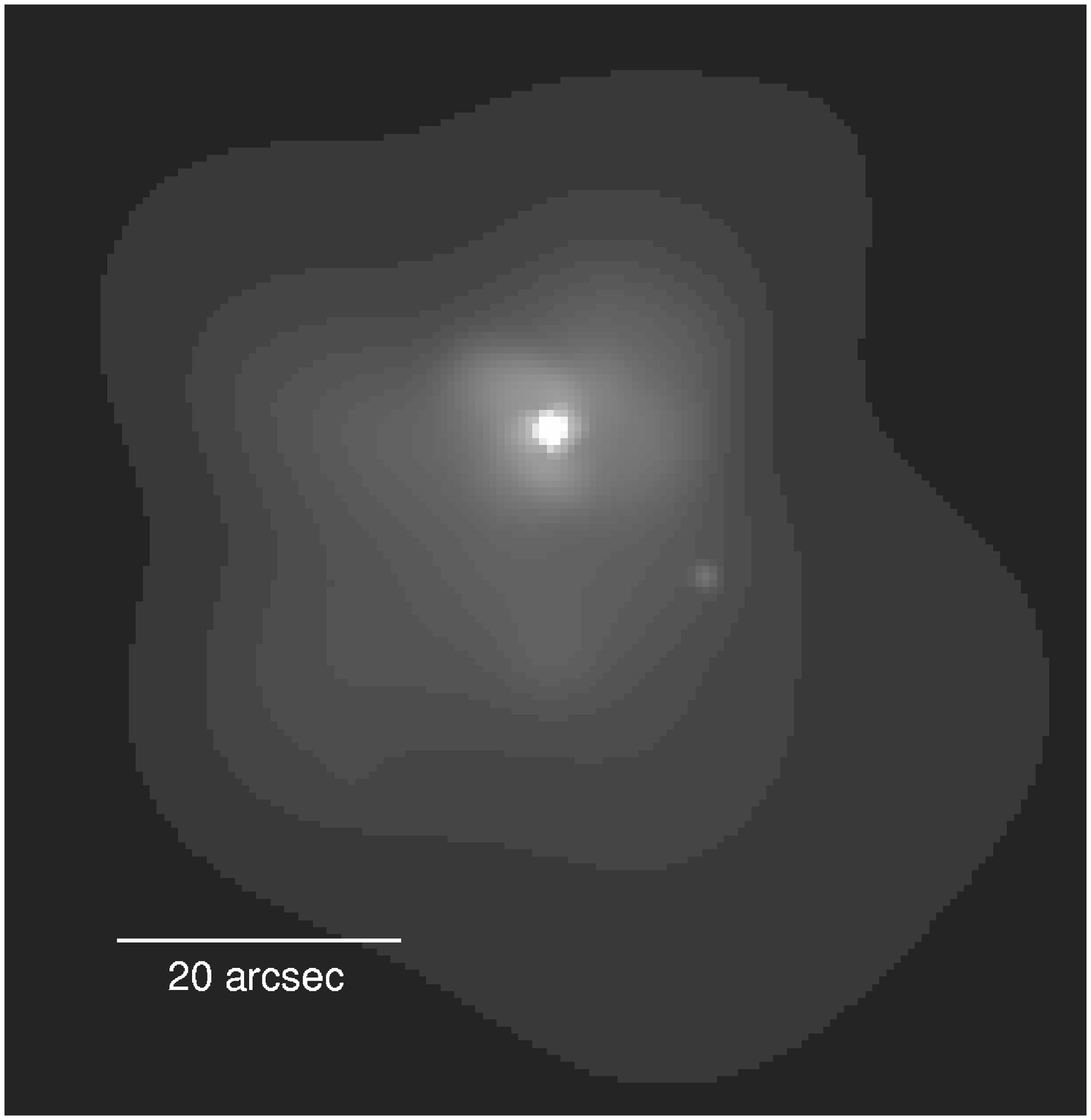}{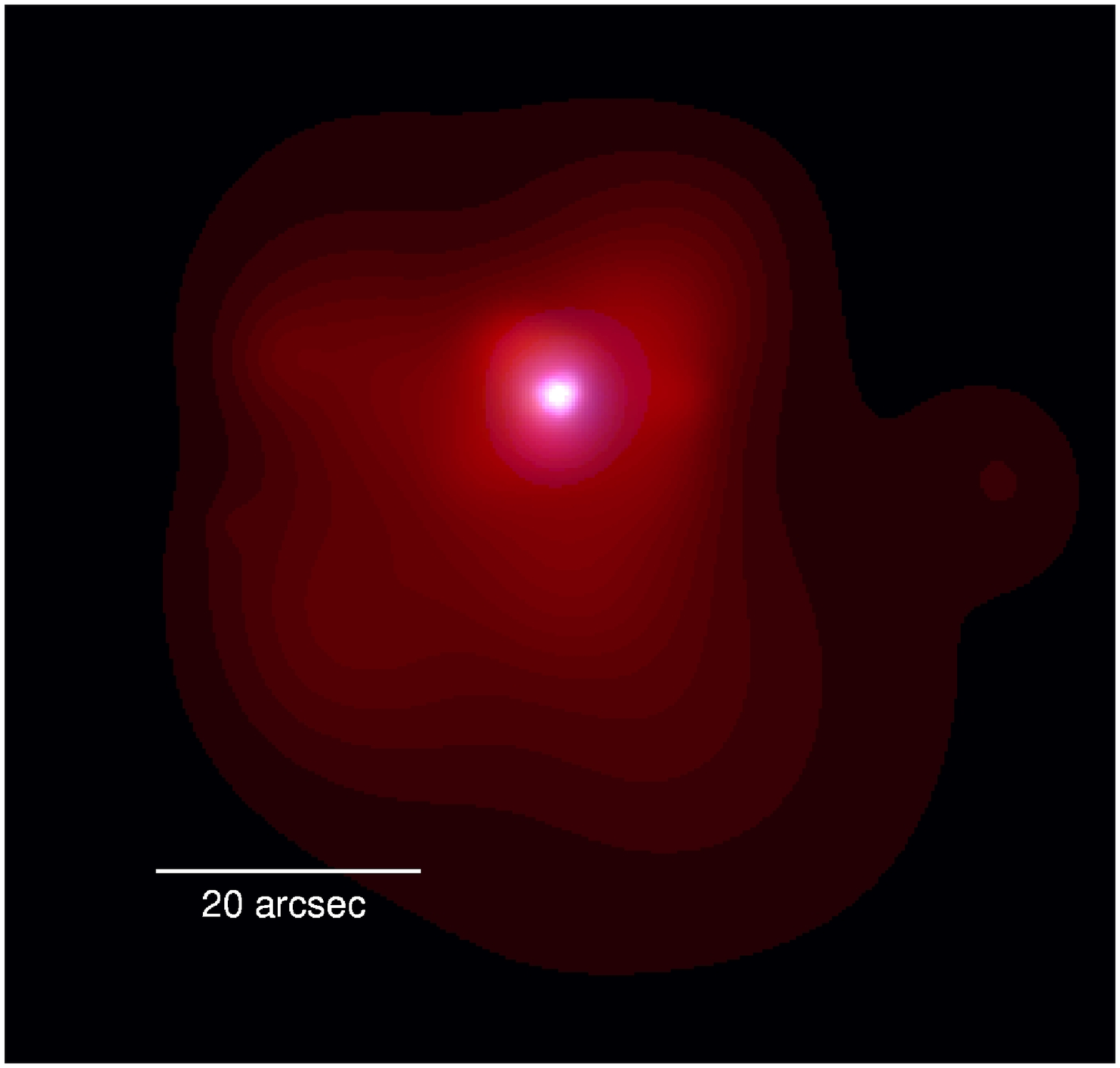}
\caption{
{\em Left panel:} Full-band, csmoothed image of \mrk\ from the merged data sets of the
three new observations.  North is up and east is to the left. 
This 114~ks exposure reveals ring structure around the nucleus as well as
a new point source to the southwest of the nucleus.
{\em Right panel:}
Three-color csmoothed image of \mrk\ from the merged data sets of
the three new observations.  Red, green, and blue correspond to
0.35--1.2, 1.2--3.0, and 3.0--8.0~keV, respectively. The blue
emission is slightly asymmetric, with an enhancement just to the west of the
nucleus.  The flat nuclear spectrum leads to the white appearance of
the nucleus.  
\label{fig:3color_im}
\label{fig:fb_im}
}
\end{figure}

\begin{figure}
\vspace{3.6in}
\includegraphics{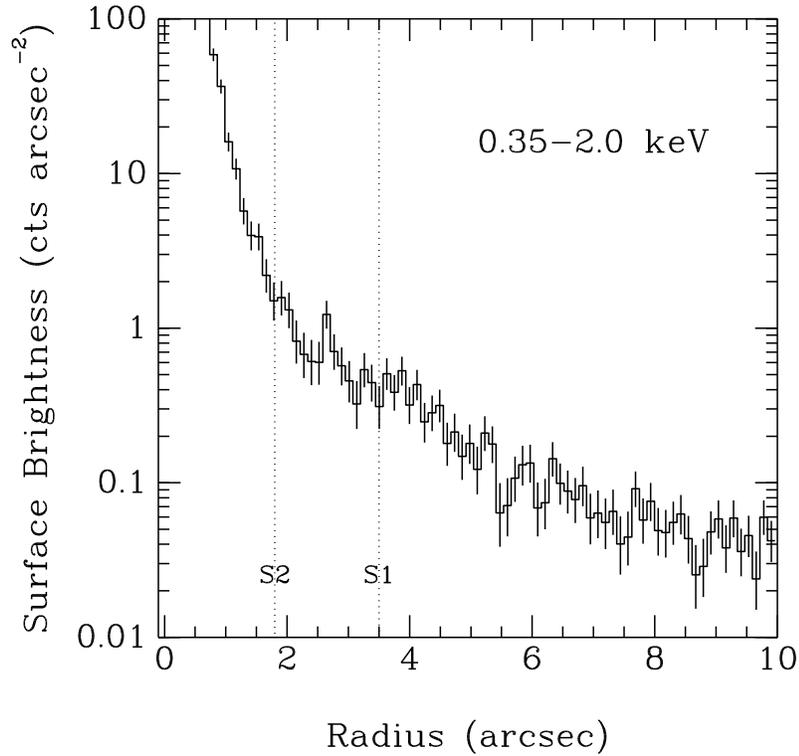}
\caption{
Radial surface brightness profile in the
0.35--2.0~keV band of the nucleus of \mrk\ from the merged data of
observations 4028--30. The 1$\sigma$ errors have been determined using
the method of \citet{Gehrels}.  The vertical, dotted lines indicate
the radii of the optical shells, S1 and S2, discussed by \citet{lipari05}.}
\label{fig:rp}
\end{figure}

From the full-band image (Figure~\ref{fig:fb_im}), the deeper data
allow us to see ring structure around the nucleus.  
A radial surface-brightness profile of the soft-band (0.35--2.0~keV)
counts centered on the nucleus (Figure~\ref{fig:rp}) reveals an enhancement
  of the X-ray emission at $R\approx2\farcs6$.  This radius is
  approximately halfway between the two bright optical `shells', S1
  and S2 (as labeled by \citealt{lipari05}), seen in \hst\ images. 
The overall galactic emission is structured and asymmetric, with X-ray
enhancements to the northeast and a
deficit to the west of the nucleus; both were previously seen in the 
1031 image.  The northeast structure is likely from photoionized (rather
than collisionally ionized) gas as it coincides with an optically bright region
attributed to \OIII\ emission \citep{HuNe1987}.  In the full-band
image, the off-nuclear point-source detected by G02 is no longer present,
however, a second one (only evident at energies $<3$~keV) 
has appeared to the southwest of the
nucleus. The inferred variability of both sources is consistent with an
identification as luminous X-ray binaries \citep[e.g.,][]{fabbiano+03}, preferentially found in
actively star-forming galaxies.

The three-color image, made by
combining 0.35--1.2 (red), 1.2--3.0 (green), and 3.0--8.0~keV (blue) csmoothed images
illustrates the energy dependence of these features (Figure~\ref{fig:3color_im}).  The asymmetry of the
hard emission surrounding the nucleus causes the image to appear bluer
to the west of the nucleus. Except for a compact region within
$\sim4\arcsec$ of the nucleus, the galaxy is not detected at energies
$>$3~keV.

\section{Optical Observations and Analysis}

\subsection{Echelle Spectroscopy}
\label{sec:het}

Mrk~231 was observed with the 9.2~m Hobby-Eberly Telescope (HET; \citealt{het_ref}) at
McDonald Observatory using the Upgraded Fiber Optic Echelle Spectrograph \citep[UFOE;][]{ufoe_ref}
in 2000 January and April. These observations were made to search for
relatively short-time scale (weeks to months) variations in the prominent
\ion{Na}{1} D absorption systems that are blueshifted 5000--6000\kms\
from the systemic velocity of the host galaxy. All of the data were acquired
using a 400~$\mu\/$m (2\arcsec) fiber with sky subtraction accomplished with a
second 2\arcsec\ fiber offset from the galaxy. The spectra have a resolution of
$\sim$0.8~\AA\ in the echelle order containing the \ion{Na}{1} D absorption
features.  As summarized in Table~\ref{tab:specpol}, the 2000 January observations consisted of
three 200~s exposures of the nucleus on January 6 followed by four 900~s
exposures on January 13. No significant differences were seen between the two
sets of observations and the results were averaged. Similarly, the 2000 April 3
dataset is an average of three 900~s exposures.

The January and April data were then inspected for differences in the
positions, widths, and equivalent widths (EWs) of the various absorption
systems of \ion{Na}{1} D. The lower spectrum shown in Figure~\ref{fig:NaD} is the
difference spectrum (Jan $-$ Apr) between the two epochs, after normalizing the
spectra to the surrounding continuum. Again, no significant differences are
seen between the two epochs, and the average of the two spectra is displayed as
the upper trace in Figure~\ref{fig:NaD}, with individual line systems
labeled according to the notation of FRM and 
quantified in Table~\ref{tab:NaD} .  At the resolution and
signal-to-noise ratio of the HET observations, the doublet splitting
is cleanly resolved and subsystems can be recognized in both systems
I and II. The D$_1$ component of system Ii is a tentative identification
within the main trough of system I at about the correct position relative to
D$_2$. Expected positions of the lines for the system III doublet \citep{BoMeMoPe1991}
are marked, but it is clear that these features were absent in early 2000.

\subsection{Ground-Based Imaging and Spectropolarimetry}
\label{sec:specpol_obs}

Optical spectropolarimetry of Mrk 231 was obtained on several nights during
2002$-$2004 with the spectropolarimeter SPOL \citep{SchStoSmi1992}.
This is the same instrument that was used for the observations published
by SSAA from 1992$-$1993, and all acquisition, calibration, and reduction
procedures are common to the two epochs.  
As with other dual-beam designs, the instrument
achieves high throughput while accurately cancelling sky background and
avoiding systematic effects to better than 1 part in 1000.  Calibration of the
polarizing efficiency of the optics is performed by observations of a
continuum source through a fully polarizing prism each observing run.
However, the use of different
interstellar polarization standard stars for the various observing runs can
result in a dispersion of up to 1\arcdeg\ in the overall registration
of the position angle. The detector was also upgraded twice in the interval between the
epochs, providing improvements in sensitivity, focus, spectral coverage, and
flat-field calibration.  The significant changes for polarimetry are an
enhancement of the signal-to-noise ratio for a given exposure time and an
improvement in spectral resolution.  As summarized in the observation log in
Table~\ref{tab:specpol}, virtually all data from the two epochs were gathered at the 2.3~m Bok
telescope of Steward Observatory using a low-resolution grating
($\Delta\lambda\sim17$\AA) with wide spectral coverage, though one dataset in
each epoch utilized a 1200 line mm$^{-1}$ grating with a resolution of 8\,\AA\ to
record better the polarization profiles of the H$\alpha$ emission and
\ion{Na}{1} D absorption features. In addition, 
observations were made at the MMT in 2004 April to investigate the diluting
effects of the host galaxy, which spans $>1\arcmin$ \citep{HaKe1987}.  Unfortunately, the utility of
these data was compromised by poor ($\sim$2\arcsec\ FWHM) seeing at the time.

We have intercompared the polarimetric results among the individual runs of
2002$-$2004 in both detail and for broad wavelength intervals such as those
listed in Table~\ref{tab:specpol}.  While the effects of variable host galaxy dilution cause
polarization changes at the level of $\Delta P=0.1-0.2\%$ (cf. the results for
the two widely different aperture sizes from consecutive nights in 2003 Feb),
we find no significant variability that can be attributed to the AGN during the
16 month period. A reinspection of the data from 1992$-$1993 leads to the same
conclusion for that period.  However, there is compelling evidence for a
decline of $\sim30$\% ($\Delta P=1-2\%$ across the bandpass) in the degree of polarization during the
decade that separates the two epochs. At the same time, the position angle
remained  constant. Means for the $\lambda\lambda6100-6600$
continuum interval are $P=3.41\%\pm0.04\%$ at $\theta=96\fdg9\pm0\fdg4$ for the
1990's versus $P=2.52\%\pm0.10\%$ at $\theta=97\fdg8\pm0\fdg7$ for the
2000's, where the uncertainties are estimated from the scatter (rms)
of the individual measurements for each epoch.
The difference is easily seen in the mean polarization spectra for the two
periods compared in Figure~\ref{fig:specpol1}, and in the binned continuum values listed in
Table~\ref{tab:specpol}.  The spectropolarimetric observations
were obtained under a variety of conditions and the emphasis was not on
accurate absolute flux calibration, but based on the general consistency in
flux level between the two epochs, any systematic change in total brightness
can be limited to $\lesssim$15\%.

Despite the change in fractional polarization level, the essential polarimetric
characteristics of \mrk\ persist between the decades.  These include a slow
rotation in position angle $\theta$ over the spectrum, significant changes in
$\theta$ between the \ion{Fe}{2} and Balmer lines versus the continuum, a
polarized flux spectrum $P \times F_\lambda$ that is bluer than the total flux
spectrum ($P_\lambda \propto \lambda^{-2}$), and emission-line features that
are much narrower in polarized versus total flux.

In the interval between the two spectropolarimetric epochs,
broad-band imaging polarimetry of \mrk\ was obtained at optical and near-UV (\S\ref{sec:impol})
wavelengths.  Ground-based $B\/$ and $R\/$ observations were made
at the Bok 2.3~m telescope
in late 1996 with SPOL in its imaging mode (a plane mirror is inserted
behind the Wollaston prism instead of the diffraction grating),
and the results are summarized in Table~\ref{tab:specpol}.
The scale of the SPOL images is 0\farcs 52\,pixel$^{-1}$.
The entries for $P\/$ and
$\theta\/$ in Table~\ref{tab:specpol} are for a 3\arcsec$\times$10\arcsec\ aperture
centered on the nucleus in the 51\arcsec$\times$51\arcsec\ images.
This rectangular aperture, oriented at a position angle of 90$^{\circ}$,
was chosen to mimic the slit and east-west extraction apertures used
for spectropolarimetry.  The optical broad-band
results for the polarization position angle are consistent with both
epochs of spectropolarimetry, and also show that
$\theta_R - \theta_B = 2$--3$^{\circ}$.  The level of polarization
in late 1996 is intermediate between the higher $P\/$ measured in the
early 1990s and the lower levels observed after 2001.  We have
searched the literature without finding accurate polarimetry of \mrk\
in this interval.  Thus, while it is likely that the polarization
change in \mrk\ was occuring during this period, we do not have data
to determine how rapidly $P\/$ varied, though the intermediate level
of the data from 1996 suggests that the change did not occur on 
very short time scales.

The effect of starlight from the host galaxy on the polarization
measurements of Mrk~231 can be estimated using the imaging polarimetry.
Figure~\ref{fig:prad} shows $P\/$ as a function of aperture radius for both filter
bandpasses. The measured polarizations at $B\/$
and $R\/$ for apertures ranging from
$\sim$1\arcsec -- 10\arcsec\ in diameter readily
show the relative increase in starlight with aperture size.
Extrapolation to $r = 0$\arcsec\
of third-order polynomial fits to the measured polarization
indicates that the intrinsic polarization of the nucleus is 6.45\% and 3.96\%
in the $B\/$ and $R\/$ bands, respectively.
The ratios of star-to-AGN light using a 10\arcsec\ aperture are
0.21 ($B\/$) and 0.64 ($R\/$). Within the typical spectroscopic extraction aperture
used for spectropolarimetry (3\arcsec$\times$10\arcsec\/),
the ratios are $\sim$0.13 ($B\/$) and $\sim$0.52 ($R\/$).

\subsection{\hst\ Imaging Polarimetry}
\label{sec:impol}
On 1998 November~28 (UT), imaging polarimetry of the inner
$\sim$1\farcs 8$\times$1\farcs 8 of Mrk~231 was obtained using the
Faint Object Camera (FOC) aboard the {\em Hubble Space Telescope}
(\hst). These observations were made in an attempt to resolve and identify the
scattering regions around the nucleus.
The f/96 optics of the FOC were used and the detector was read out in its
smallest format (128$\times$128~pixel) to mitigate saturation problems from
the bright nucleus.
To minimize the contribution of host-galaxy starlight, 
the observation was made in the near ultraviolet with the F346M filter
($\lambda_0 = 3480$~\AA ; $\Delta\lambda_{\rm FWHM} = 434$~\AA ).
This choice of filter also took advantage of the high polarization
of the nucleus ($P \sim 15$\% at $\lambda \sim 3000$~\AA ) measured by SSAA.

Four exposures ranging from 1166~s to 1363~s were made with each of three
Polaroid filters in the beam combined with F346M.
About 1.4~hr was spent using each of the POL60 and POL120
Polaroid filters, while $\sim$1.5~hr total exposure time was used for POL0.
In general, care must be taken with FOC polarimetry because the throughputs of the
polarizer filters are not identical; however, in the F346M bandpass
they are quite similar. In addition, the instrumental polarization
introduced by off-axis optics inserted during the 1993 Space Shuttle
\hst\ servicing mission  to correct the telescope's spherical aberration 
is not well determined \citep{capetti96}.
Given the high polarization of Mrk~231, we have ignored the 1--3\%
instrumental polarization in the analysis of the FOC data since any
resolved polarized regions will have even higher polarization than
the $\sim$15\% observed in the near UV for the integrated AGN light
(SSAA).
The polarimetry was reduced in the manner described by
\citet{capetti95}.

The imaging data do not show any evidence for extended structure.
Identification of faint extended emission very close to the point source
is made especially difficult because each
Polaroid filter yields a distinct point spread function (PSF).
No PSF calibration images exist for the individual filter
combinations, and so
accurate subtraction of the PSFs is not possible.
However, aperture polarimetry of the images supports the conclusion that at
the resolution of the FOC, Mrk~231 is unresolved in the near UV.
Circular apertures with radii of 3~pix (0\farcs 043) to 33~pix (0\farcs 469)
all yield polarizations of $P \approx 15.5\%$ at $\theta \approx 99^{\circ}$.
The measurement of $P\/$ is close to the value found at $\sim$3000~\AA\ in
a much larger aperture by SSAA,
though the measured position angles
are discrepant by nearly 10$^{\circ}$.
The difference could be caused by the instrumental polarization of
the FOC. 
Also, the same polarization values are found for the faint
emission surrounding the core of the PSF. Aperture polarimetry using an annulus
with an inner radius of 6~pixels from the centroid of the bright core and an
outer radius of 33~pixels also yields $P \approx 15\%$.
The lack of a significant difference in the polarization
between the core of the PSF and the much fainter flux surrounding the core
suggests that the flux in the wings of the PSF completely dominates any
resolved emission sources in the nuclear region.

The FOC data are not useful for tracking the variations in $P\/$ because
no other observations cover this wavelength region.
What we can conclude from the near-UV imaging and polarimetry is that the
nucleus and the regions scattering AGN light are small.
We set a conservative upper limit for the extent of the
scattering regions by using the worst of the FOC PSFs, when POL60
was in the telescope beam, with FWHM$=3.1$~pixel.
The data are therefore consistent with the near-UV light of the AGN (both polarized and
unpolarized sources) being contained in an unresolved point source with a
FWHM$\simeq 0\farcs044$. For \mrk, this corresponds to a radius of
19~pc.

Past attempts to resolve the nucleus of Mrk~231
at various wavelengths have
also yielded only upper limits to the size.
Ground-based near-IR and 12.5~$\mu$m observations, as well as early \hst\
measurements using the Near-Infrared Camera and Multi-Object Spectrometer
(NICMOS) at 1.6~$\mu$m have set limits of $\sim$100~pc for the FWHM of
the unresolved nucleus \citep{LaEtal1998,SoEtal2000,quillen01}.
Evidence for a rotating molecular torus located 30--100~pc away from
central continuum source is inferred by the observations of hydroxyl maser
emission \citep*{klock03}. 
At 18~cm, \citet{lonsdale03}
detected emission extending 0\farcs 03
southward of an unresolved core component with a $\sim$4~pc FWHM.
Recently, archival NICMOS data obtained in 1998 were re-analyzed,
taking full advantage of the stable PSF of the instrument \citep*{low04}.
They set an upper limit to the diameter of the core of $<$7~pc at 1.1~$\mu$m.
As mentioned above, the sophisticated PSF analysis employed by \citet{low04} cannot be
used for the FOC data because of the large differences in the PSFs
observed. 

\section{Results and Discussion}

\subsection{X-ray Continuum and \feka\ Emission}

With the direct X-ray continuum of \mrk\ blocked by a Compton-thick
absorber \citep{braito+04}, the $>3$~keV \chandra\ and \xmm\ nuclear
spectra are composites of indirect X-rays that reach the observer
through reflection off of neutral material and/or scattering by
ionized plasma into the line of sight. 
In this scenario, the \feka\ emission is expected to be spectrally
unresolved at S3 CCD resolution (FWHM$\sim$200~eV at 6.4~keV). 
Intrinsically broad \feka\ emission is
generated in or near the inner accretion disk; such a line would be
completely obscured by the same absorber that blocks the direct continuum.  
Our interpretation of the broad \feka\ feature
detected with \xmm\ by TK as a composite of narrow lines is more
consistent with the combined \chandra\ data (with better
photon statistics; see Fig.~\ref{fig:xspec}) as well as the physical
understanding of the system.  The best-fitting energies for two narrow
lines fit separately in the \xmm\ and combined \chandra\ spectra are 6.3 and
6.6~keV, suggesting reflection off neutral or nearly neutral material (to
generate the lower energy fluorescent emission) and scattering off
more highly ionized plasma (or neutral plasma with an outflow velocity
of $\sim10,000$\kms).  
This is generically consistent with the preferred model of
G02 with multiple scattered lines of sight to the nucleus comprising
the composite \mbox{2--8~keV} spectra.  However, at present the relationship
between the \feka\ emission lines and the continuum emission is unclear.

A closer examination of the individual \chandra\ spectra
indicates that two narrow \feka\ lines are not detected in all four
observations, and therefore these features could be flux-variable on timescales of weeks.
However, this is speculative at present given the photon statistics of
the individual spectra.  

We confirm the 2--8~keV continuum variability of $\sim45\%$ within 6 hrs
presented in G02 with a reanalysis of the \chandra\ data from observation 1031 
using archival data and an updated version of the CIAO
software.  Variability is not detected within any of the
new observations, and the overall 3--8~keV flux variability between
the 4 \chandra\ observations is $\lesssim15\%$.  The difference between the 3--8~keV
flux of the \xmm\ observation (2001 Jun) and the combined \chandra\
observations (2003 Feb) is $\sim$+20\% over 20 months.  This is
significantly larger than the measured 2--10\,keV flux difference between ACIS-S3 and
EPIC-pn of $\sim5\%$ \citep{snowden}, and therefore likely indicates
actual long term variability in the hard-band flux.  

\subsection{\ion{Na}{1} D Absorption-Line Variations}

The absorption lines involving \ion{Na}{1} D were organized by 
\citet{BoEtal1977}
into redshift systems I (observed wavelength 6050\,\AA) and II
(6015\,\AA), and a system at 5984\,\AA\ was added by 
\citet{BoMeMoPe1991}.
Today, the latter group is generally known as system III, though it was termed system
IV as recently as \citet{KoDiHa1992}.\footnote{The original system III of \citet{BoEtal1977} 
appears only in the Balmer and \ion{Ca}{2} H,K lines and is
taken to indicate the systemic redshift of the host galaxy.} All three
\ion{Na}{1} D systems are apparent in the SPOL spectrum of SSAA from
1992$-$1993 and in our 2002$-$2004 spectrum (both shown at an expanded scale in
Figure~\ref{fig:specpol2}).  System IIa remained approximately unchanged during the decade
spanned by the observations, but the system I trough appears to have weakened
slightly and system III nearly disappeared altogether. These effects are not
simply a result of a difference in spectral resolution, as the spectra were
obtained with the identical grating (resolution 8\,\AA), and other features in
the same spectra, such as the narrow peak atop H$\alpha$ evident in Figure~\ref{fig:specpol2},
are as apparent in the 2002$-$2004 data.

\begin{figure}
\vspace{6.0in}
\includegraphics{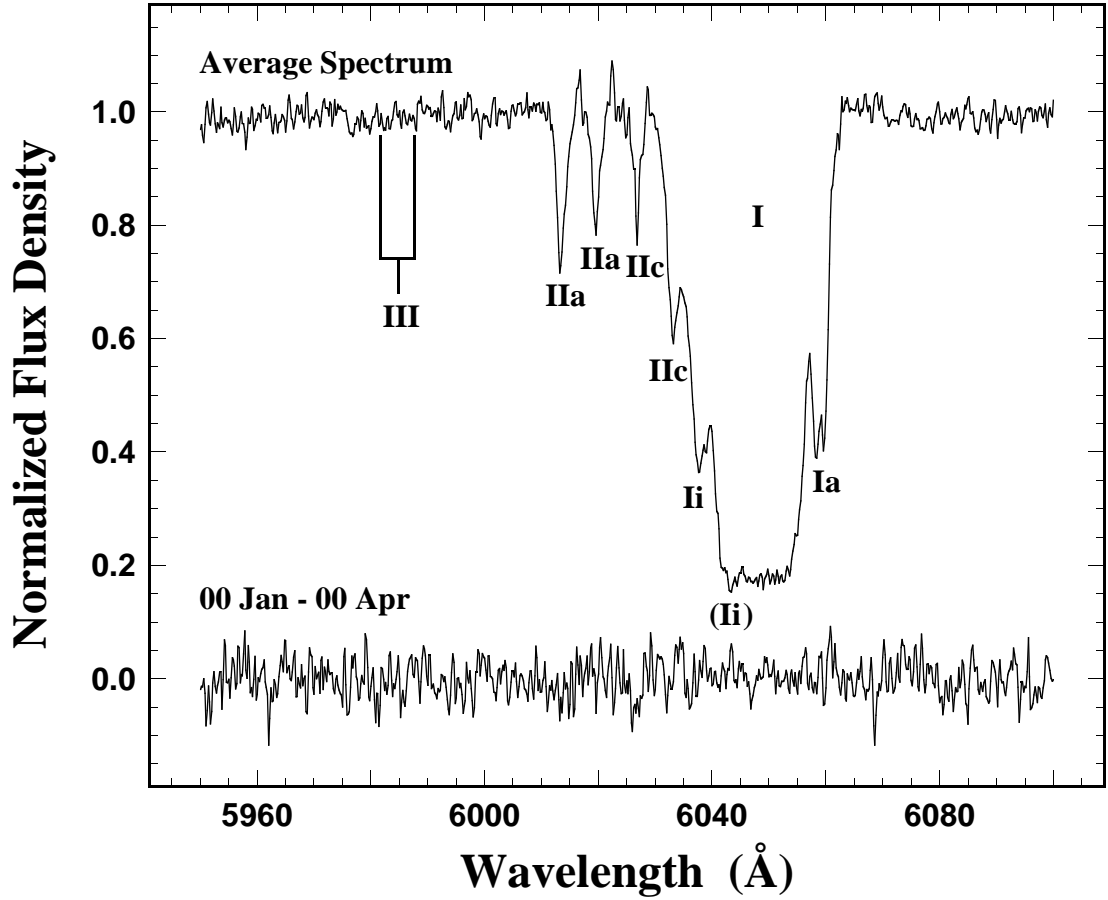}
\caption{
The \ion{Na}{1} D absorption systems of Mrk~231 during 2000.  The
average spectrum of 2000 January and April is shown with individual line
systems labeled.  Systems I and II were first identified by 
\citet{BoEtal1977},
and are seen to break into subsystems Ia, Ii, and IIc in these much
higher resolution HET echelle observations (see also FRM).
The D$_1$ component of system Ii is tentatively identified within
the main trough of system I at about the correct position relative to D$_2$.
Positions of the system III lines are marked, but they are not
detected in these observations of early 2000. Shown below the average spectrum is the difference
spectrum (Jan $-$ Apr), which shows little evidence for change in the $\sim$4
month interval.
\label{fig:NaD}
}
\end{figure}

\begin{figure}
\vspace{8.5in}
\includegraphics{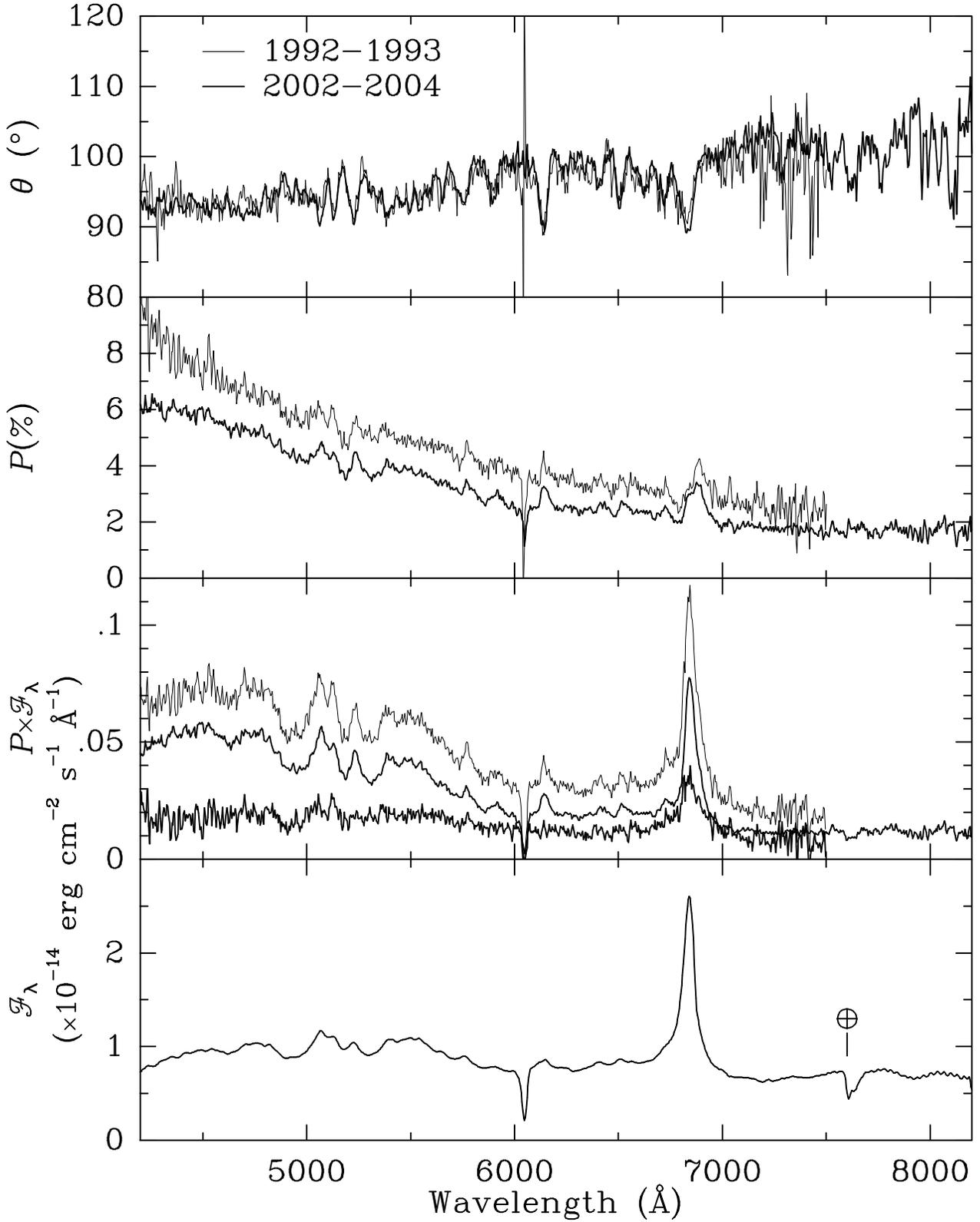}
\caption{Comparison of the observed-frame, co-added polarization spectra of Mrk 231 from
2002$-$2004 {\em (thick line)} and 1992$-$1993 {\em (thin line)}.  Note the systematic
decline by nearly 30\% ($\Delta P=1-2\%$ across the bandpass) in the
degree of polarization with largely unchanged position angles.  
The polarized flux panel ($P\times
F_\lambda$) also shows the difference spectrum between the two epochs (1990's
minus 2000's).  
A representative total flux spectrum from 2003 (bottom
panel) is shown for comparison.  
Note the much narrower profiles of the \ion{Fe}{2} and
\Halpha\ emission lines in the polarized flux spectrum versus the
total flux spectrum.  
\label{fig:specpol1}
}
\epsscale{1.0}
\end{figure}

\begin{figure}
\vspace{8.5in}
\includegraphics{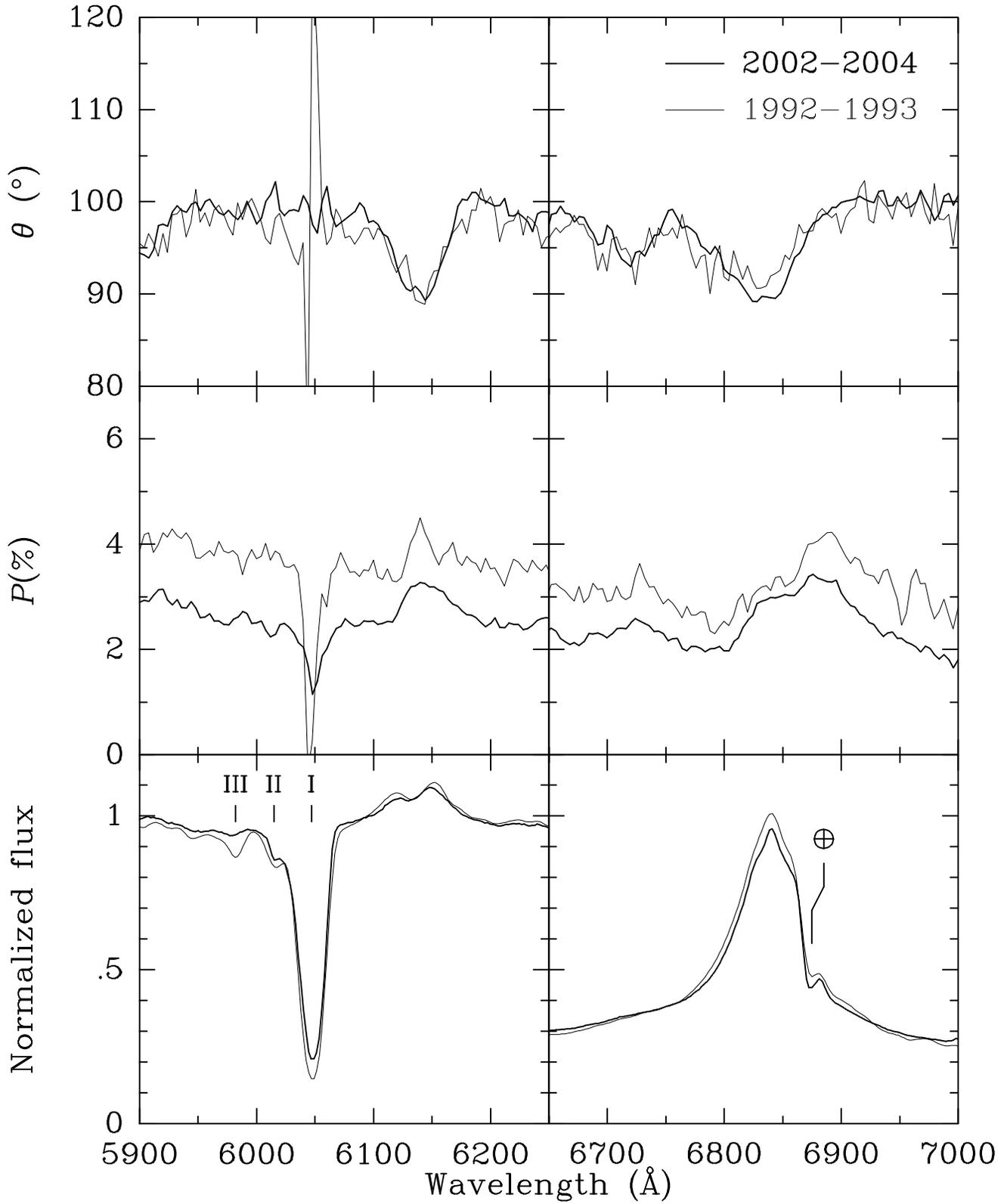}
\caption{As in Figure~\ref{fig:specpol1} for the spectral regions surrounding 
the \ion{Na}{1} D and
H$\alpha$ lines. Roman numeral labels in the \ion{Na}{1} D panel indicate
the absorption systems described by \citet{BoEtal1977} and \citet{BoMeMoPe1991}.
\label{fig:specpol2}
}
\end{figure}

\begin{figure}
\vspace{3.7in}
\includegraphics{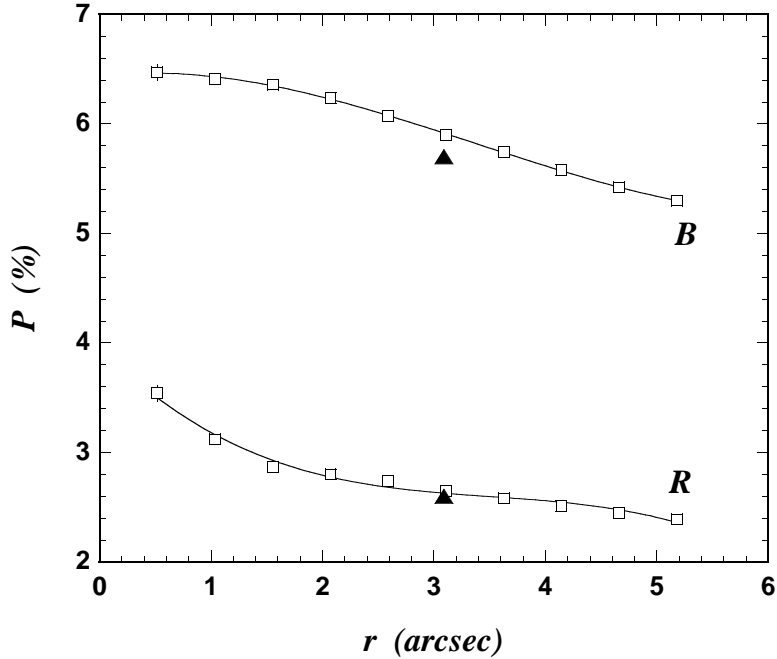}
\caption{Aperture polarimetry of Mrk~231 on 1996 December~7 at
$B\/$ and $R\/$.  In both colors there is a clear trend for $P\/$
to decrease as the size of the measurement aperture is increased, 
indicating the inclusion of more unpolarized light from the host galaxy out to 
$\sim$10\arcsec\ from the unresolved AGN.  {\it Triangles\/} represent the
polarization with a 3\arcsec$\times$10\arcsec, east-west aperture that
approximates a typical spectropolarimetric extraction aperture used for
Mrk~231.
\label{fig:prad}
}
\epsscale{1.0}
\end{figure}

The rise and fall of the system~III \ion{Na}{1} D lines have been well
chronicled (e.g., \citealt{KoDiHa1992}; FRM).  The most recent
characterization by \citet{lipari05}
features a smooth decrease in line
strength from a maximum (EW = 2.2\,\AA) in 1988 July to EW = 0.36\,\AA\ in the
data of \citet{rupke02}
obtained in 2001 February. However, data were
lacking for a 7 yr interval of the decline.  Our HET UFOE spectra
shown in Figure~\ref{fig:NaD} were obtained near the end of the hiatus, yet neither D$_1$
nor D$_2$ were detected to an individual upper limit of EW $\sim$ 0.03\,\AA,
well below the detection of \citet{rupke02}.  At the same time, our spectra from the
spectropolarimeter in 2002$-$2004 confirm their detection at approximately
the same strength (Figure~\ref{fig:specpol2}).  The inference is that system III essentially
disappeared altogether for a period commencing at some point after the observations of FRM in
1994 April, was absent during 2000 January and April, and reappeared weakly by
early 2001.  \citet{rupke02} 
also claim that the observed redshift of the
feature appears to be increasing with time, though much of the historical data
do not resolve the doublet.

Multi-component decomposition of the \ion{Na}{1} D complex has been carried out
under various assumptions (\citealt{RuStFo1985}; FRM; \citealt{rupke02}),
and will not be repeated here.  We simply note that, apart from the absence of system
III, the HET spectra closely resemble the data of
\citet{rupke02} 
obtained one year later. The higher resolution of the UFOE data allows a
closer examination of the narrow features near the edges of the principal
system I feature.  The IIc doublet, whose D$_1$ component lies at the blue
edge of the main system I trough, has progressively increased in strength over
the past decade.\footnote{System IIb has not been seen since the 1991 spectrum
of FRM.}  In fact, the now-obvious D$_2$ component lies in what was a region
of the continuum in the 1991 data of FRM and was only beginning to appear in
their 1994 spectrum.  There is marginal evidence that system IIc D$_2$ is
slightly narrower than the system IIa features, as indicated in Table~\ref{tab:NaD},
however it is clear that the assignment of lines to distinct systems is
meaningful only for bookkeeping purposes. There is also marginal evidence that
system Ia D$_1$ and Ii D$_2$ are each composed of two narrow components with
a separation of $\sim$1~\AA.\ We consider the identification of the Ii D$_1$
component at the blue edge of the main system~I trough to be tentative, and a
higher signal-to-noise ratio spectrum would be useful for separating components in this
region.  Finally, we point out a curious excess in the ``continuum''
immediately surrounding the system IIa D$_1$ line in Figure~\ref{fig:NaD}.  A similar
feature is apparent at somewhat lower signal-to-noise ratio in the 1994 spectrum of FRM;
together the observations suggest the existence of a narrow emission component
of \ion{Na}{1} D associated with one of the absorption systems (see also FRM
and \citealt{lipari94} on this issue).

As pointed out by FRM, an interpretation of the main system I trough in terms
of a few highly saturated doublets \citep{RuStFo1985}
is not feasible because
a column density in neutral Na as high as $10^{15}$~cm$^{-2}$ develops strong
damping wings, inconsistent with the observed profile.  FRM's choice of a
number of optically thin components ($N_{\rm Na}\sim10^{13}$~cm$^{-2}$) over
several moderately saturated doublets is based on the presence of residual
flux at the bottom of the trough.  However, they appear to have ignored the
underlying galaxy starlight, which our imaging has shown is significant in the
typical observing aperture of a ground-based telescope. The general flatness
of the system~I trough suggests that the individual components are optically
thick, so it would seem that a blend of a number of moderately saturated lines
is still an option, and the individual system column densities may well be intermediate
between the above two estimates.

Starlight as the sole source of residual light in the system I trough is
consistent with the lack of polarization in the bottom of the trough in
1992$-$1993 (Figure~\ref{fig:specpol2}), and SSAA estimated the galactic dilution of the AGN
polarization accordingly.  From Figure~\ref{fig:NaD}, we also find that the residual flux
level in the echelle spectrum of the trough, $\sim$17\% of the surrounding
continuum level, is consistent with the host galaxy dilution expected within a
2\arcsec\ aperture.  In contrast, the 2002$-$2004 spectropolarimetry indicates
measurable polarization in the trough, with a position angle common to that
of the surrounding continuum.  To be sure, the net polarization in this
feature is a factor of two below that of the surrounding continuum, even
considering the decreased continuum polarization measured in 2002$-$2004.
However, its presence suggests that some scattered light may have been
introduced into the trough by 2002. A small amount of scattered AGN light that
avoids absorption in the main system I trough would tend to support our
tentative identification of the Ii D$_1$ feature above.

A quantitative estimate of the intrinsic degree of polarization of the
scattered light in the system~I trough of Mrk 231 is strongly affected by the
uncertainty in the host-galaxy contribution within the observing aperture, but
it could exceed the polarization of the surrounding continuum.  In
high-$z$ BAL quasars, scattered light completely dominates the residual flux in
the rest-frame UV troughs, and the measured polarization can exceed 10\% \citep{SH99,ogle99}.

\subsection{Nature of the Optical Polarization Change}

The magnitude of the polarization observed in Mrk~231 and the similarity of
characteristics between the emission line and continuum polarization have long
been interpreted as indicating an origin in scattering 
\citep{Thompson+80,SchMil85}.
In red, strongly polarized quasars in general, and
Mrk~231 in particular, there is compelling evidence in the shapes of the
polarized and total flux spectra that the scattering occurs in extensive dust
clouds located within $\sim$100 pc of the nucleus 
(\citealt{GoMi1994}; SSAA; \citealt{hines+01,Smith+03,Smith+04})
Again specific
to Mrk~231, our \hst\ imaging polarimetry presented above places these
clouds interior to a 20~pc radius, and the fact that polarization position
angle structure is observed across the H$\alpha$ emission profile (\S\ref{sec:specpol_obs};
SSAA) indicates that the broad-line region is partially resolved from the
vantage point of the scattering medium, i.e., much of the
scattering could occur
far deeper in the nucleus.  The 1.1\,$\mu$m nuclear light constrained by 
\citet{low04}
to originate inside a radius of $\sim$3~pc is apparently emitted by
hot dust grains, the survival of which is possible as close as  $\lesssim$1~pc
to a source with $L\sim4\times10^{12}$~$L_\sun$.  In view of these
facts, we imagine that {\em all} of the optical emission escaping from the nucleus
may have undergone at least one scattering.  Indeed, with optical extinction
in interstellar clouds scaling as $N_{\rm H}/A_V \sim 2\times10^{21}$~cm$^{-2}$
mag$^{-1}$ \citep[e.g.,][]{BH78},
the very high absorption measured toward the X-ray power law 
\citep[$N_H \sim 2\times10^{24}$\cmsq;][]{braito+04}
would suggest that there are no open sightlines to the central source
from soft-X-ray to submillimeter wavelengths.  
However, as noted by \citet{maiolino+01}, the column densities from X-ray and
optical extinction measures in AGN often imply much lower dust-to-gas ratios
than typical interstellar clouds.  Furthermore in \mrk, the X-ray
absorber may be largely dust-free; it is likely to
be much more compact than both the longer wavelength continuum emission
and the dusty shroud obscuring the optical--UV continuum source.

Because scattering in a clumpy dust shroud
is prone to the searchlight effect (redirection of the illuminating
beam), it is not possible to relate directly
the time scale for a polarization change to motion of clouds.  However, the
orbital period for gas at a distance of 0.5~pc from a 10$^9$~$M_\sun$
black hole is 1000~yr, and so a significant change in scattering geometry over a
decade is not implausible.

The following essential phenomena must be explained with regard to the observed
polarization change in Mrk 231: 1) a decline in degree of polarization
$P$, and in polarized flux $P \times F_\lambda$ by a factor of $\sim$1/3
(e.g., $P_{\rm obs}=3\%$ to 2\% at $\lambda\sim7000$\,\AA) in both the
continuum and the emission lines; 2) a position angle of polarization that has
considerable structure across the spectrum but is manifestly constant with
time; 3) an upper limit on the amount of any systematic change in total
nuclear brightness of 15\%; 4) a spectrum of polarized flux that has
narrower emission lines than that in total flux; and 5) a reduction in
the depth of the system I trough as well as the appearance in 2002 of
significant polarization in this feature at the systemic polarization position
angle.

\subsubsection{The Importance of Starlight Dilution}

An increase in the relative dilution by an unpolarized component is an obvious
means of decreasing both the degree of polarization, $P$, and the
polarized flux, 
$P \times F_\lambda$, without modifying the position angle. As an example, a
decline in the intrinsic brightness of the AGN, while the diluting host galaxy
starlight remains constant, would lead to a drop in the polarization measured
within a finite aperture.  The allowed magnitude of the decline is, of course,
subject to our conclusion that the overall nuclear brightness varied by
no more than 15\% during the decade.

The lack of an aperture dependence to the polarization measurements summarized
in Table~\ref{tab:specpol} already argues that starlight cannot be a strong diluting component
to the light measured through a typical spectropolarimetric aperture and
thus is not an important factor in the polarization change. The
strongest evidence for this conclusion is provided by the residual flux in the
base of the system I \ion{Na}{1} D trough, measured at 15\% in 1993 and 21\%
in 2003 in the 3\arcsec\ aperture used with the spectropolarimeter.  With
stars initially accounting for no more of 15\% of the total light at 6000\,\AA,
a change of nearly 1/3 in polarized flux at that wavelength requires the AGN
to have faded by more than a factor of four.  The net spectrum in 2002 would
have been far fainter than what was measured and the resulting \ion{Na}{1} D
trough would have been $<$60\% of the continuum level.

Additional evidence of a change in polarization of the AGN
component itself comes from the fact that the intrinsic (AGN-only) $B$-band
polarization determined from the imaging polarimetry in 1996 ($P_B=6.45\%$,
Figure~\ref{fig:prad}) is no larger than the {\it net\/} (AGN $+$ galaxy) value computed from
the spectropolarimetry for the same filter passband in 1992$-$1993 (6.62\%).
Indeed, the imaging polarimetry implies that the $B$-band polarization
appropriate to a 3\arcsec$\times$10\arcsec\ aperture in 1996, 5.70\%
(Table~\ref{tab:specpol}), 
is already well on its way to the 2002$-$2004 value of 5.04\%.

Finally, we note that increased dilution by the opening of a new, direct
(i.e., unscattered) sightline to the AGN could avoid the problem with the
depth of the \ion{Na}{1}\,D 
trough if the direct beam also passed through the system I
clouds. However, this mechanism would leave the polarized flux unchanged, contrary
to what is observed.

\subsubsection{A Strongly Polarized Component}

An important clue toward a more successful model for the polarization change
was noted in \S\ref{sec:specpol_obs}: namely, the spectrum of polarized flux is not simply
a reflection of the total flux spectrum. Specifically, the individual \ion{Fe}{2}
and \ion{H}{1} emission features are much more sharply defined in the former.
This is not simply a matter of contrast, but more of line width, and implies
that the scattering clouds do not see the full velocity dispersion of the
broad line region.  The physical origin of the difference will be addressed below, but here
we point out that the difference between the two spectra {\em requires} that a
large fraction of the AGN total flux spectrum, containing the continuum,
\ion{Na}{1} D absorption, and smeared emission lines, has no polarization
whatsoever. This leads to an upper limit on the amount of light that becomes
polarized by scattering (because scattering is not 100\% efficient at
inducing polarization), and therefore to a lower limit on the degree of
polarization of the scattered light.

We have carried out simple experiments to estimate the fraction of the total
flux spectrum from 2002$-$2004 that is scattered by simply summing the
spectrum of $P \times F_\lambda$ and $F_\lambda$ in various relative fractions
and judging by eye when the sharper features of the former are no longer
visible in the total. Conservatively, we find that a higher degree of structure
would be apparent in the total flux spectrum if a spectrum like that of the
polarized flux comprised more than 30\% of the total at a wavelength of
5500\,\AA.  The implications are that the more sharply structured, polarized AGN
component must be diluted by a virtually unpolarized AGN component (and starlight) to
a ratio of at least 2.3:1, and that the intrinsic degree of polarization of
the scattered light at this wavelength in 2002 was a substantial $P_{\rm sca} \ge
3.3 \times P_{\rm obs} = 12.5\%$!  This is comparable to the observed
polarization level at blue optical and near-UV wavelengths (SSAA), suggesting
that the unpolarized AGN component is primarily responsible for the overall
very reddened optical-UV spectral energy distribution, and that its diluting
effects are greatly reduced at short wavelengths because of dust
extinction. Because the polarized
component is such a small fraction of the total light at visible
wavelengths, a drop in the net
observed polarization from 4.8\% to 3.8\% between the 1990's and 2000's at
5500\AA\,\ only requires a decline in brightness of the polarized AGN component
of $<$30\%.  	     
The change in brightness of the overall nucleus (AGN $+$ starlight)
is then $<$14\%, consistent with the limit on the brightness change between the
epochs.

\ \ \ \ \ ~
\ \ \ \ \ ~

\ \ \ \ \ ~
\ \ \ \ \ ~

\subsubsection{More Complex Mechanisms}
Alternate explanations for the polarization change in \mrk\ can be
constructed by the introduction of additional polarized components between the
epochs.  We have investigated this option by attempting to synthesize the
observed flux and polarization spectra for the two epochs using the method
applied by \citet{SchHinSwi02} to the bipolar protoplanetary nebula
surrounding CIT 6.  We find that such a model can reproduce a large amount of
the polarization structure seen in the Mrk~231 spectrum, and that the observed
changes can be interpreted as a change in the relative brightness of the
polarized components.  For example, the opening of a new scattered sightline
with a polarized flux spectrum like the difference spectrum shown in Figure~\ref{fig:specpol1},
but with a polarization position angle orthogonal to the systemic angle, i.e.,
$\theta\sim10^{\circ}$, provides an excellent match to the most recent
observations.  If the fractional polarization of the new sightline were large,
$P\gtrsim10\%$ at 6500\,\AA, the overall increase in brightness through our
spectroscopic aperture would be $<$15\%.  At some level, changes in $\theta$
between epochs are also expected from this interpretation. Unfortunately, as
discussed by \citet{SchHinSwi02} and SSAA, even simple two-component models
suffer from being non-unique, with a range of combinations of polarization,
position angle, and flux level reproducing the observations with equal quality.

\subsection{Dust-Reddened Quasars and the Structure of AGN}

Mrk~231 can be considered the first known example of a population of optically
red, IR-luminous AGN with quasar-like total energy outputs that are now being
identified in significant numbers by near-IR, optical and
X-ray surveys (e.g., \citealt{hines+01,Smith+02b,matt02,zakamska+05}
and references therein). These objects are found in both Type~1 (broad
permitted lines) and Type~2 (narrow permitted lines) varieties, and are distinguished from other
samples of quasars by the fact that a sizeable fraction of each type
($\gtrsim$10\%) show optical polarization of the total light $P\ge3\%$, the
traditional threshold for ``highly'' polarized AGN. Generally, the polarized
Type~2 quasars have been shown to follow the basic predictions of the Unified
Scheme for AGN \citep{antonucci93} by revealing broad permitted lines
characteristic of a Type~1 object in polarized light. This indicates the
existence of a broad line region and attests to anisotropic obscuration
surrounding the central direct continuum source \citep[e.g.,][]{zakamska+05}.  The
optical obscuring structure is nominally regarded as a dusty/molecular torus at
$r\approx1$--100~pc \citep[e.g.,][]{nenkova02}, which is seen at high
inclination (i.e., edge-on) in the
Type~2 systems, and the polarization is produced by scattering off particles
located near the polar axis.  

Though the Unified Scheme has been very effective in accounting for many key
aspects of the various empirical classes of Seyfert-luminosity AGN within a single simple
framework, it provides no obvious explanation for the strongly polarized Type~1
Seyfert galaxies and quasars like Mrk~231, which by virtue of the strong
broad-line component in their total flux spectra are thought to be viewed at
smaller inclination angles --- i.e., through the ``hole'' of the torus.  In total flux,
the new population of red quasars shows evidence for considerable
reddening, whereas the polarized flux spectra are considerably bluer \citep{Smith+03}. 

In an attempt to account for the $\sim$1\% polarization often seen in Type~1
AGNs, \citet{GoMi1994} and more recently \citet{jsmith+02,jsmith+05}
proposed the existence of an equatorial disk nominally coplanar
with and lying inside the torus. Scattering off dust and/or electrons in this
disk provides an additional polarizing mechanism for Type~1 objects. Depending
on inclination angle, the observed polarization of a Type~1 AGN might be
dominated by scattering off particles near the polar axis (the
predominant Type~2 polarizing mechanism, for $i\sim45^\circ$) or off the equatorial disk (for
$i\lesssim45^\circ$). The scheme of \citet{jsmith+05} is currently a geometrical
framework rather than a physical model, but there are clear parallels with
disk-launched winds \citep{KK94,MuChGrVo1995}.  If the equatorial
scattering medium is identified with the disk wind, then this mechanism
should not be an important source of polarization in BAL quasars such
as \mrk.

\subsection{Toward a Model for the Nucleus of Mrk 231}

Based on the multiwavelength data available on \mrk, the nucleus is
clearly complex with more than one structure scattering continuum and
emission-line photons into the line of sight. Insight into the nucleus on the 
smallest physical scales in \mrk\ is provided by X-ray
observations.  The most interesting constraints are provided by the
combination of X-ray data from different observatories.  The
confirmation of Compton-thick absorption from the detection of the
direct continuum at hard X-rays by \citet{braito+04} coupled
with the continuum variability detected by G02 point towards
a compact Compton-thick absorber.  This opaque absorber must lie
interior to the broad-line region, and therefore would not be identified
with a molecular torus as are the Compton-thick absorbers in Type~2
AGNs.  The X-ray absorber is likely to be on scales
smaller than or comparable to the UV-optical continuum region,
to enable detection of the continuum and radiative driving of the BAL
outflow. As discussed in G02, the variability of indirect 2--8~keV X-rays constrains the scale
of the scatterer to be $\sim10^{15}$\cmsq\ and consequently the
ionization of this plasma is expected to be high (i.e., with Fe almost
totally stripped of electrons).  The new combined \chandra\ data indicate
that highly ionized plasma may have been detected in emission (with one \feka\
 emission component at $\sim6.6$~keV), 
though the relationship between the \feka-emitting gas and
the observed 2--8~keV continuum emission is unclear.  The dominant
scattering mechanism in this case is electron scattering, and so this
structure is distinct from the dust-enshrouded scatterer dominating the
optical polarization.

\begin{figure}
\vspace{4.0in}
\includegraphics{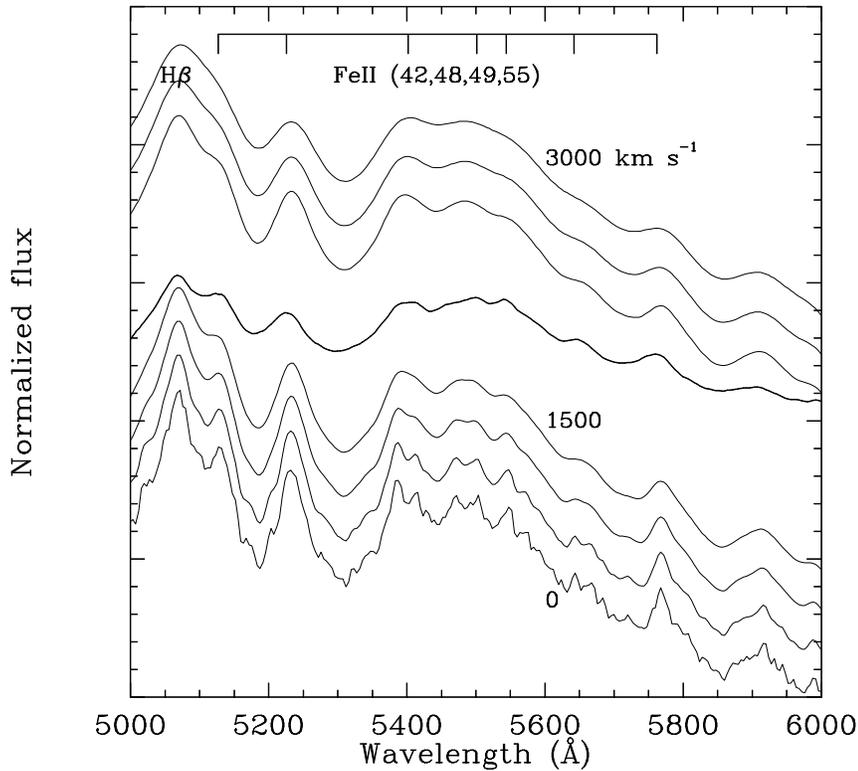}
\caption{The polarized flux spectrum of Mrk~231 for 5000--6000~\AA\ in the
observed frame (thin curves) convolved with gaussians ranging from no line
broadening (the observed spectrum; bottom) to FWHM = 3000~km~s$^{-1}$ (top
spectrum) in increments of 500~km~s$^{-1}$. The thick curve is the observed
total flux spectrum with the starlight contribution of the host galaxy within
the observing aperture subtracted. Broadening the observed polarized flux
spectrum by 1500$-$2000~km~s$^{-1}$ yields an emission-line sharpness similar
to that of the total light spectrum.
\label{fig:specpol3}
}
\end{figure}

On larger scales, two polarization characteristics suggest
that there exists a prevailing symmetry in the unresolved active nucleus, and
that some of the structures inferred by the Unified Scheme may be relevant.
First, the constancy of the polarization position angle over 3 decades,
despite a sizable change in $P$, requires that the dominant optical scattering
region(s) maintain the same orientation relative to the continuum source and
broad line region. Second, the
individual \ion{Fe}{2} and \ion{H}{1} emission features are much more sharply
defined in the polarized flux spectrum than in the total flux spectrum. The
difference is better displayed in Figure~\ref{fig:specpol3}, where we compare the total flux
(less the contribution of the host galaxy; thick line) and polarized flux
spectra (thin lines) for the observed-frame 5000$-$6000\,\AA\ region, with
various amounts of gaussian smoothing applied to the polarized flux.  As can
be seen, the spectrum of $P \times F_\lambda$ must be convolved with a
gaussian of FWHM = 1500$-$2000\kms\ to approximate the smoothness of
emission features in the total flux spectrum. Also apparent in Figure~\ref{fig:specpol3} is
the lack of perfect correspondence between the velocity centroid of features in the
two spectra, with discrepancies as large as $\sim$500\kms\ evident
between the two. Some of these displacements are likely associated with the
small polarization position angle rotations observed across the profiles of
prominent emission features.

An attractive explanation for the narrower spectral features in the polarized
flux spectrum is that the scattering plane is essentially 
orthogonal to the direction of an
important dynamical motion of the line-emitting gas.  Scattering off
particles in polar lobes, as in the Unified Scheme for AGNs, offers such an
arrangement, if the broad-line gas orbits the nucleus in a disk whose axis is
nominally coincident with the lobes.  The lobes also provide a natural
explanation for the very large polarization inferred for this component in
\S\ref{sec:specpol_obs}. The broad line region disk would also be
roughly coplanar with the dust torus that presumably determines
whether we perceive a Type~1 or 2 object in the total flux
spectrum. Since Mrk~231 is a Type~1 AGN, the dust torus is not fully edge on, 
but it still serves to define the polar scattering region(s) at a position
angle of $\sim$7$^\circ$.  The polarization properties are also consistent with
an equatorial BAL outflow, though the covering fraction of the BAL
wind is apparently greater than that of the torus.
Furthermore, the detection of some polarized
continuum unabsorbed by \ion{Na}{1}\,D (Fig.\,\ref{fig:NaD}) shows that the wind does not
cover the entire sky.

The polarization position angle is closely aligned with the axis of the 40-pc
scale triple source (core-lobe) imaged at GHz frequencies 
(\citealt{UlWrCa1999} and references therein). 
On parsec scales, a double radio source in the nucleus is misaligned by
$\sim65^{\circ}$ from the larger structure \citep{ulvestad99b}, and
therefore not coincident with the position angle of the scattering medium.
Ignoring any velocity dispersion within the
scattering lobes, differential projection effects across the lobes, and the
unknown inclination of the orbiting disk to the line of sight, a velocity of
1500$-$2000~km~s$^{-1}$ corresponds to gas orbiting 1$-$2~pc away for
a black hole at the high end of the mass distribution, 
10$^9~M_\odot\/$.  
Even this upper estimate for the
radius is well within the upper limit to the AGN
size determined by \citet{low04}.
A broad line region with dimensions of a few pc is
larger than the $\sim$0.3 pc estimated for objects of comparable
luminosity ($\lambda L_{{\rm 5100\AA}}\sim10^{45}$\lumin)
using the reverberation technique \citep{kaspi+00}, but the assumptions
made above suggest that the velocity dispersion estimate is a lower
limit, and so the disk size may be overestimated.

\citet{Smith+04} also consider the specific case of Mrk~231 and, based on
the redness of the optical continuum, the presence of BAL features, and the
general constancy of the polarization position angle across the spectrum,
conclude that scattering in polar lobes is the dominant polarizing mechanism.
However, their assignment to ``far-field'' scattering in the lobes is not
supported by our much higher signal-to-noise ratio data that reveal considerable position angle
structure across the Balmer and \ion{Fe}{2} emission lines, nor by the
\hst\  polarimetry result that locates the scattering material no further than 20~pc from
the continuum source.  In fact, the results of \citet{low04} suggest that
the scattering in the polar lobes likely takes place within just a few pc of
the nucleus; a much more compact scattering region than typically assumed in
the Unified Scheme or actually observed in several nearby Seyfert galaxies and
some quasars \citep[see e.g.,][]{zakamska+05} that show scattered light from
regions of the order $\gtrsim$1~kpc. However, light polarized by
scattering from such a compact region would likely also be blocked by
the same structure that blocks the broad-line region (with similar
scales) in Type~2 objects.

The observations also require that the
compact primary scattering region be enshrouded in dust to account for
the lack of a narrow emission-line region. In this view of Mrk~231, the much
reduced size of its polar scattering regions and the fact that light from
these regions must also traverse a significant amount of dust are significant
departures from the simple unified picture of AGNs. The compactness of the
scattering regions and the ubiquitous dust distribution around the nucleus and
possibly within the polar lobes may, in fact, be responsible for the large
decrease in $P\/$ observed between the early 1990s and early 2000s. One
possible scenario is that dusty material interior to the main scattering
clouds changed the amount of illumination within the scattering lobes or,
alternatively, dust exterior to the polar lobes blocked more of the scattered
light along our line of sight. In either case, large flux changes are not
necessary to cause the polarization variation, and the position angle of the
polarized flux will be preserved. Finally, the augmented morphology proposed
by \citet{Smith+04}, i.e., an additional scattering disk coplanar with the
dust torus, but interior to it, may be responsible for the variation in
polarization position angle across emission features and the overall rotation
of $\theta\/$ observed throughout the optical and UV (SSAA).

\subsection{Future Investigations}

The holistic examination of the indirect X-ray, UV, and optical
emission has provided some new insights into the nuclear
structure of \mrk\ that bear further investigation.
From the close examination of the individual \chandra\ spectra, 
it appears that two narrow \feka\ lines are not equally strong in all four
observations. A claim of flux variability between the \feka\
 lines is speculative at present given the photon statistics of
the individual spectra. With its higher effective area at $E>3$~keV
compared to \chandra,  additional $\sim30$~ks \xmm\ observations with similar time sampling
to these \chandra\ observations would be sensitive to this
phenomenon.  A detection of \feka\ line variability on these time scales
would constrain the location of scattering material to within the
broad line region, as suggested by the continuum variability detected
by G02 (and assuming the scattering media are also the sources of the
\feka\ emission).  A single, long-look (100--300\,ks) \xmm\ observation would
also be productive for searching for additional hard-band variability
on intermediate time scales of $\sim$day, for studying the iron-band
features in depth, and for potentially detecting 
soft-band emission lines from the starburst and AGN with the
Reflection Grating Spectrometer.

The lack of \OIII\ emission in the optical spectrum indicates
that the covering fraction of dust around the nucleus is very high,
thereby preventing ionizing photons from illuminating low-density
gas in what would otherwise be the narrow-line region.  
The large-column density X-ray absorber clearly does not
cover the entire sky, otherwise short-timescale variability would not
be detectable. As X-rays can easily penetrate dusty gas that would block FUV
photons, the X-ray narrow-line region (ionized by X-ray-energy
photons) should exist and may even have already been
detected (TK).  In this region, the strongest features expected are the forbidden
lines of \mbox{\ion{O}{7}$\lambda$22.1\,\AA} and \mbox{\ion{Ne}{9}$\lambda$13.7\,\AA} \citep[e.g.,][]{ogle03}. 
High-resolution X-ray spectroscopy with the next
generation of X-ray observatories such as \conx\ will enable a detailed
emission-line study undiluted by the strong direct continuum, 
a rare opportunity to study the X-ray narrow-line region of a
nearby, quasar-luminosity AGN.  

The polarization variation we have detected in Mrk~231 is perhaps the most
accurately recorded so far among radio-quiet AGN, but it is not unique.  I Zw
1, which also exhibits a polarized spectrum caused by scattering, displayed a
reduction in fractional amount and a rotation of $\sim$50$^\circ$ in position
angle between the 1980's and 1990's \citep{Smith+97}.  
\citet{GoMi1994} 
noted a 25\% decline in the polarization of Mrk~509 over a 2 yr period, and a
similar variation in the same object was discussed by \citet{young+99}. 
\citet{jsmith+02} also point out apparently significant changes in Akn~120
and NGC~7469 over periods of $\sim$18 months.  The fact that all are Type~1
AGNs suggests that a significant fraction of the polarization
of these objects arises within the inner few pc of the nucleus, and offers hope
that further studies of polarization and its variability might continue to
improve our understanding of their nuclear structure.

\acknowledgements{We thank Jane Turner for generously providing
  reduced \xmm\ spectra for comparative analysis, and Joan Wrobel and
  Jim Ulvestad for helpful discussions. Financial support was provided
  by NASA through {\em Chandra} grants GO3--4137 and GO3--4137B, and
  {\em HST} grant GO--6444. Support for SCG was
  provided by NASA through the {\em Spitzer} Fellowship Program, under
  award 1256317.}


\clearpage
\begin{landscape}
\begin{deluxetable*}{lcccc}
\tablecolumns{5}
\tablewidth{4.5in}
\tablecaption{\chandra\ Observing Log
\label{tab:xlog}}
\tablehead{
\colhead{Obs. ID} &
\colhead{Date} &
\colhead{Exposure} &
\multicolumn{2}{c}{Nuclear Count Rates\tablenotemark{a} ($10^{-2}$ ct~\persec)}\\
\colhead{} &
\colhead{} &
\colhead{(ks)} &
\colhead{0.5--2.0 keV} &
\colhead{2.0--8.0 keV} 
}
\startdata
1031   &  2000 Oct 19 &   39.3 &  $1.98\pm0.07$ &   $1.92\pm0.07$ \\
4028   &  2003 Feb 03 &   39.7 &  $1.86\pm0.07$ &   $1.98\pm0.07$ \\
4029   &  2003 Feb 11 &   38.6 &  $1.83\pm0.07$ &   $2.00\pm0.07$ \\
4030   &  2003 Feb 20 &   36.0 &  $1.79\pm0.07$ &   $2.04\pm0.08$ \\
\enddata
\tablenotetext{a}{The values are background-subtracted, though the
  background was negligible in the hard band and $\lesssim3\%$ in the
  soft band.   Nuclear source counts were extracted from a
  $3\arcsec$-radius source cell. Background counts were collected from
  an annulus with inner and outer radii of $5\arcsec$ and 
  $10\arcsec$, respectively, and normalized to the area of the source cell.}
\end{deluxetable*}
\begin{deluxetable*}{lccccccc}
\tablecolumns{8}
\tablewidth{0pt}
\tablecaption{\chandra\ X-ray Spectral Fitting\tablenotemark{a} \label{tab:xspec}}
\tablehead{
\colhead{} &
\multicolumn{7}{c}{Observation Identification} \\
\colhead{Parameter (Units)} &
\colhead{1031} &
\colhead{4028} &
\colhead{4029} &
\colhead{4030} &
\colhead{4028--30} &
\colhead{\xmm\tablenotemark{b}} &
\colhead{\xmm\tablenotemark{c}} 
}
\startdata
3--8~keV Count Rate ($10^{-2}$ \persec) &
1.32$\pm0.06$ &
1.38$\pm0.06$ &
1.52$\pm0.06$ &
1.52$\pm0.07$ &
1.51$\pm0.04$ &
$\cdots$ &
$\cdots$ \\
5.8--6.7~keV Count Rate ($10^{-2}$ \persec) &
0.21$\pm0.02$ &
0.18$\pm0.02$ &
0.23$\pm0.03$ &
0.19$\pm0.02$ &
0.19$\pm0.01$ &
$\cdots$ &
$\cdots$ \\
$F_{\rm 3-8}$ ($10^{-13}$\,erg\,cm$^{-2}$\,s$^{-1}$) &
4.65$^{+0.29}_{-0.35}$ &
4.80$^{+0.35}_{-0.34}$ &
5.17$^{+0.35}_{-0.35}$ &
5.05$^{+0.37}_{-0.37}$ &
5.03$^{+0.20}_{-0.20}$ &
4.16$^{+0.26}_{-0.27}$ &
4.16$^{+0.27}_{-0.26}$ \\
$L_{\rm 3-8}$  ($10^{42}$\,erg\,s$^{-1}$) &
1.83$^{+0.09}_{-0.14}$ &
1.87$^{+0.14}_{-0.13}$ &
2.03$^{+0.14}_{-0.14}$ &
1.98$^{+0.15}_{-0.15}$ & 
1.97$^{+0.08}_{-0.09}$ & 
1.60$^{+0.10}_{-0.10}$ &
1.66$^{+0.10}_{-0.11}$ \\
$\Gamma$      &
0.47$^{+0.82}_{-0.42}$ &
0.48$^{+0.28}_{-0.47}$ &
0.32$^{+0.40}_{-0.32}$ &
0.51$^{+0.24}_{-0.49}$ &
0.41$^{+0.20}_{-0.24}$ &
0.83$^{+0.35}_{-0.35}$ &
0.76$^{+0.36}_{-0.29}$ \\
Power Law Normalization ($10^{-5}$\,$f_{\rm E}$\tablenotemark{d}) &
2.20$^{+4.56}_{-1.10}$ &
2.20$^{+1.35}_{-1.01}$ &
2.09$^{+1.63}_{-0.80}$ &
2.88$^{+1.22}_{-1.53}$ &
2.31$^{+0.79}_{-0.68}$ &
3.71$^{+2.56}_{-2.20}$ &
3.45$^{+2.60}_{-1.27}$ \\
Line Energy (keV) &
6.34$^{+>0.66}_{->0.34}$ &
6.68$^{+0.08}_{-0.07}$ &
6.34$^{+0.04}_{-0.04}$ &
6.4               &
6.31$^{+0.08}_{-0.11}$/6.64$^{+0.05}_{-0.07}$ &
6.47$^{+0.13}_{-0.13}$ &
6.31$^{+0.11}_{-0.11}$/6.62$^{+0.10}_{-0.12}$ \\
$\sigma$ (keV) &
0.45$^{+0.72}_{-0.41}$  &
0.01  &
0.01  &
0.01  & 
0.01/0.01 &
0.19$^{+0.12}_{-0.10}$ &
0.01/0.01 \\
Line Flux ($10^{-6}$\,phot\,cm$^{-2}$\,s$^{-1}$) &
6.00$^{+16.0}_{-4.7}$ &
2.81$^{+2.03}_{-1.70}$ &
3.09$^{+2.13}_{-1.62}$ &
$<1.86$~~~&
0.93$^{+0.73}_{-0.84}$/1.25$^{+0.97}_{-0.81}$ &
3.56$^{+2.21}_{-1.81}$ &
1.65$^{+1.02}_{-1.09}$/1.73$^{+1.11}_{-1.09}$ \\
Line EW (eV) &
638$^{+1700}_{-499}$ &
281$^{+204}_{-170}$ &
264$^{+183}_{-139}$ &
$<160$\,~~~&
70$^{+62}_{-55}$/98$^{+77}_{-64}$&
483$^{+283}_{-279}$ &
143$^{+88}_{-95}$/167$^{+108}_{-105}$ \\
Edge Energy (keV) &
7.12 &
7.12 &
7.12 &
7.12 &
7.12 &
7.12 &
7.12 \\
$\tau_{\rm max}$  &
0.29$^{+0.71}_{-0.29}$ &
0.11$^{+0.76}_{-0.11}$ &
0.49$^{+0.70}_{-0.49}$ &
0.17$^{+0.75}_{-0.17}$ &
0.28$^{+0.31}_{-0.27}$ &
0.23$^{+0.31}_{-0.23}$ & 
0.28$^{+0.27}_{-0.28}$ \\
$\chi^2/\nu$       &
23.2/27 &
22.4/29 &
21.7/31 &
13.1/31 &
60.1/72 &
23.0/27 &
21.6/26 \\
$P(\chi^2|\nu)$    & 
0.68 &
0.80 &
0.89 &
0.99 &
0.65  &
0.67 &
0.71 \\
\enddata
\tablenotetext{a}{Errors are for 90$\%$ confidence taking 1 parameter
  to be of interest ($\Delta\chi^2=2.71$). Values with no errors were
fixed for the fitting.  The errors for $F_{\rm 3-8}$, the
observed-frame 3--8~keV flux, and $L_{\rm 3-8}$, the rest-frame 3--8~keV luminosity, were
determined from $90\%$ confidence errors in the power law
normalization with all other parameters frozen to the best-fit values.}
\tablenotetext{b}{The \xmm\ EPIC-pn data from 3.0--10.0~keV were fit for this
  analysis. These spectral-fitting results are from the same data
  set presented in TK, and are fully consistent.}
\tablenotetext{c}{The same data, fit with two unresolved Gaussians
  instead of a single Gaussian with $\sigma$ free to vary.}
\tablenotetext{d}{Units of $f_{\rm E}$ are phot\,cm$^{-2}$\,s$^{-1}$\,keV$^{-1}$.}
\end{deluxetable*}
\clearpage
\end{landscape}
\begin{deluxetable}{cccccccc}
\tablewidth{4.5in}
\tablecaption{Log of Optical and UV Observations \label{tab:specpol}}
\tablehead{
\colhead{Date}
& \colhead{Exposure}
& \colhead{Coverage}
& \colhead{$\Delta\lambda$}
& \colhead{Aperture}
& \colhead{Telescope}
& \colhead{$P$\tablenotemark{a}}
& \colhead{$\theta$\tablenotemark{a}}
\\
\colhead{(yyyymmdd)}
& \colhead{(s)}
& \colhead{(\AA)}
& \colhead{(\AA)}
& \colhead{(\arcsec)}
&
& \colhead{(\%)}
& \colhead{($^\circ$)}}
\startdata
\multicolumn{8}{c}{Spectropolarimetry}\\
\tableline
\multicolumn{8}{l}{First Epoch (1990's)}\\
19920312 & ~~\,800 & 4200$-$7900 & 17 & 3\by10 & Bok   & 3.36 & 97.5 \\
19930226 & ~\,1200 & 4100$-$7400 & 20 & 4\by8~ & Bok   & 3.41 & 96.9 \\
19930428 & ~\,1600 & 4000$-$7300 & 17 & 3\by8~ & Bok   & 3.40 & 96.6 \\
19930513 & ~\,4800 & 5630$-$7360 & ~\,8 & 3\by9~ & Bok   & 3.46 & 96.6 \\
\multicolumn{6}{l}{Mean values{\dots}} & 3.41 (0.04) & 96.9 (0.4) \\
\tableline
\multicolumn{8}{l}{Second Epoch (2000's)}\\
20021204 & ~\,2880 & 4200$-$8200 & 17 & 3\by10 & Bok   & 2.45 & 96.7 \\
20030202 & ~\,4800 & 4200$-$8200 & 17 & 3\by7~ & Bok   & 2.65 & 98.1 \\
20030203 & ~\,2400 & 4200$-$8200 & 20 & 4\by14 & Bok   & 2.42 & 97.3 \\
20030203 & 12800 & 5250$-$7700 & ~\,8 & 3\by13 & Bok   & 2.33 & 97.0 \\
20030307 & ~\,3200 & 4200$-$8200 & 17 & 3\by10 & Bok   & 2.57 & 97.3 \\
20040426 & ~~\,960 & 4200$-$8400 & 16 & 0.7\by5~~~\, & MMT & 2.55 & 98.3 \\
20040426 & ~~\,960 & 4200$-$8400 & 19 & 1.1\by5~~~\, & MMT & 2.55 & 98.2 \\
20040426 & ~~\,960 & 4200$-$8400 & 22 & 1.4\by5~~~\, & MMT & 2.56 & 98.4 \\
20040426 & ~~\,960 & 4200$-$8400 & 24 & 1.8\by6~~~\, & MMT & 2.56 & 98.8 \\
\multicolumn{6}{l}{Mean values{\dots}} & 2.52 (0.10) & 97.8 (0.7)\\
\tableline
\multicolumn{8}{c}{Imaging Polarimetry}\\
\tableline
19961207 & ~~\,960 & $B$ & $\sim$1000 & 3\by10 & Bok & 5.70 & 93.4 \\
19961207 & ~~\,320 & $R$ & $\sim$1600 & 3\by10 & Bok & 2.60 & 95.9 \\
19981128 & 15520 & F346M & $\sim$430 & 0.47 & \hst\ & 15.3 & 97.8 \\
\tableline
\multicolumn{8}{c}{Echelle Spectroscopy}\\
\tableline
20000106 & 3$\times$200 & 4534$-$9154 & 0.8 & 2 (fiber) & HET & \nodata & \nodata \\
20000113 & 4$\times$900 & 4534$-$9154 & 0.8 & 2 (fiber) & HET & \nodata & \nodata \\
20000403 & 3$\times$900 & 4416$-$8810 & 0.8 & 2 (fiber) & HET & \nodata & \nodata \\
\enddata
\tablenotetext{a}{For the spectropolarimetric observations, the mean polarization
properties are computed over the observed range 6100$-$6600\,\AA, a
largely continuum band common to all of the datasets.  For a single dataset,
uncertainties are dominated by calibration, seeing, and other 
aperture-dependent effects as indicated
by the standard deviations (in parentheses; $\Delta P \sim 0.1$\%,
$\Delta\theta \sim 1^\circ$). See \S\ref{sec:specpol_obs} for further discussion.}
\end{deluxetable}
\begin{deluxetable}{lccrrrr}
\tablecolumns{7}
\tablewidth{4.5in}
\tablecaption{\ion{Na}{1} D absorption in Mrk~231
\label{tab:NaD}
}
\tablehead{
\colhead{System\tablenotemark{a}}  &
\colhead{$\lambda_0\/$} & \colhead{$\lambda$\tablenotemark{b}} &
\colhead{$v\/$} &
\colhead{$v - v_{\rm sys}$}\tablenotemark{c} &
\colhead{FWHM} &
\colhead{EW} \\
\colhead{} &
\colhead{(\AA )} &
\colhead{(\AA )} &
\colhead{(\kms)} &
\colhead{(\kms)} &
\colhead{(\kms)} &
\colhead{(\AA )}}
\startdata
I & 5889.950+5895.923 & 6047 & 7750\ \ \ \  & $-$4650 & $\sim 1100$ & $\sim 18$\ \ \ \ \ \ \ \\
Ia & 5895.923 & $6058.82 \pm 0.05$ & $8168 \pm 3$ & $-$4232 & $110 \pm 8$ & $0.63 \pm 0.12$ \\
Ib & 5889.950 & $6037.97 \pm 0.07$ & $7440 \pm 3$ & $-$4960 & $101 \pm 8$ & $0.34 \pm 0.11$ \\
 & & & & & & \\
IIa & 5889.950 & $6013.49 \pm 0.03$ & $6222 \pm 2$ & $-$6178 & $92 \pm 6$ & $0.54 \pm 0.02$ \\
IIa & 5895.923 & $6019.60 \pm 0.03$ & $6223 \pm 2$ & $-$6177 & $81 \pm 8$ & $0.34 \pm 0.02$ \\
IIc & 5889.950 & $6026.81 \pm 0.10$ & $6885 \pm 5$ & $-$5515 & $73 \pm 9$ & $0.28 \pm 0.03$ \\
IIc & 5895.923 & $6033.28 \pm 0.12$ & $6903 \pm 6$ & $-$5497 & $48 \pm 15$ & $0.13 \pm 0.04$ \\
\enddata
\tablenotetext{a}{The system identifications are adopted from FRM.}
\tablenotetext{b}{Uncertainties were estimated from Monte Carlo
simulations using the noise measured in the continuum.}
\tablenotetext{c}{The measured radial velocity of the line system
relative to the systemic redshift of 12,400\kms.}
\end{deluxetable}

\end{document}